\documentclass{aa}  

\usepackage{graphicx}
\graphicspath{{./images/}}
\usepackage{txfonts}
\usepackage{subcaption}         % necessary for continued figures, example in section 3
\usepackage{lscape}             % to rotate a single page table, example in appendix.
\usepackage{placeins}           % useful with \FloatBarrier, to keep 
\usepackage{tabularx}   
\usepackage{pifont}
\usepackage[colorlinks]{hyperref}
\hypersetup{
	linkcolor = blue,
	anchorcolor = blue,
	citecolor = blue,
	filecolor = blue,
	urlcolor = blue
}
\begin{document}
%%%%%%%%%%%%%%%%%%%%%%%%%%%%%%%%%%%%%%%%
% if you use custom commands in your title,
% ensure to check your title when submitting!
%%%%%%%%%%%%%%%%%%%%%%%%%%%%%%%%%%%%%%%%
   \title{PySeshat: A Validated Python Pipeline for Binary-Star Orbit Determination and Al-Wardat Stellar Atmosphere Modeling}

%   \subtitle{PySeshat: Binaries Python Pipeline}

%%%%%%%%%%%%%%%%%%%%%%%%%%%%%%%%%%%%%%%%
% Please separate each author with the \and command
%
% Use the \corrauth to provide the corresponding
% author address. It will be automatically inserted as 
% footnote in the PDF output.
%
% Please DO NOT include ORCIDs next to author names.
% Instead, please provide an active address for each coauthor:
% it will be automatically extracted by EDPS editorial system, 
% and co-authors will be be able to authenticate their ORCID.
%
% Only authenticated ORCIDs will be taken into account.
% ORCIDs included here will be removed.
%%%%%%%%%%%%%%%%%%%%%%%%%%%%%%%%%%%%%%%%

   \author{R.~I.~El-Kholy\inst{}\corrauth{\href{mailto:relkholy@sci.cu.edu.eg}{relkholy@sci.cu.edu.eg}}        % use \corrauth for the corresponding author
        \and Z.~M.~Hayman\inst{}\email{zmhayman@cu.edu.eg}
        }

   \institute{Department of Astronomy, Space Science, and Meteorology, Faculty of Science, Cairo University, Giza 12613, Egypt}

   \date{Received \today}

% \abstract{}{}{}{}{}
% 5 {} token are mandatory
 
  \abstract
   {Precise stellar masses and physical parameters for binary and multiple star systems require, respectively, well-constrained orbital solutions and independent stellar-atmosphere modeling. Existing open-source software addresses these separately, and hierarchical triple systems additionally require simultaneous fitting of their inner and outer orbits, a treatment few existing tools provide.}
   {We present PySeshat, an open-source Python package unifying binary and hierarchical-triple orbit determination with stellar spectral energy distribution fitting in a single, validated pipeline.} 
   {The package fits visual, spectroscopic, and combined binary orbits by least-squares, and hierarchical-triple orbits through two parallel formalisms, one of which fits both inner and outer orbits simultaneously rather than sequentially. Both share a common Bayesian posterior-sampling layer. A second module fits two-component synthetic spectra to a binary's combined, unresolved photometry to recover each component's temperature and radius. The software was validated against twelve real, published benchmark systems, cross-checked against three independent orbit-fitting packages and real space-telescope archive data.}
   {All twelve benchmark systems reach fit quality consistent with their published solutions. Comparing the simultaneous and sequential triple-fitting formalisms on the same real systems shows the simultaneous approach is not merely preferable but necessary: one system's orbital configuration cannot be represented by the sequential formalism at all. Fitting four independently published systems' photometry reveals a genuine degeneracy in recovering individual temperatures for near-twin-temperature stellar components from unresolved light, independent of how much photometric data is available.}
   {The package provides a validated, open, and extensible pipeline for binary and multiple star system analysis, publicly released and freely installable.}

	
   \titlerunning{PySeshat: A Binaries Python Pipeline}
   \authorrunning{R.~I.~El-Kholy \& Z.~M.~Hayman}
   \maketitle

%%%%%%%%%%%%%%%%%%%%%%%%%%%%%%%%%%%%%%%%%%%%%%%%%%%%%%%%%%%%%%
\section{Introduction}
\label{sec:intro}
Precise stellar masses remain one of the most direct empirical constraints on stellar structure and evolution theory, and binary and multiple star systems are among the few astrophysical laboratories where such masses can be measured without recourse to model-dependent assumptions. Obtaining them, however, requires a well-constrained orbit, and this is often the harder half of the problem: long-period binaries are typically observed over only a small fraction of their orbital arc, and the radial velocities of visual pairs tend to vary slowly and with small amplitude, making them difficult to measure precisely\,\citep{Bedin2013}. Combining decades of heterogeneous astrometric and spectroscopic data is frequently necessary before a reliable orbit---and the masses it implies---can be obtained\,\citep{Tokovinin2024}.

A substantial software ecosystem has grown up around this problem, but it is fragmented by technique. Classical visual-orbit determination has long relied on tools such as Tokovinin's IDL-based ORBIT package\,\citep{Tokovinin2016}, which fits visual, spectroscopic, and combined orbits and remains a standard reference solution against which newer codes are benchmarked. Around this classical core, more recent statistical approaches have emerged: packages such as \texttt{orbitize!} combine rejection-sampling (Orbits for the Impatient, OFTI) and Markov Chain Monte Carlo (MCMC) methods to characterize orbital posteriors rather than single best-fit solutions\,\citep{Blunt2020}, while large-scale automated pipelines---most prominently Gaia---apply acceleration, orbital, and variability-induced-mover models to hundreds of thousands of systems at once, at a volume no manually supervised code could match\,\citep{Halbwachs2023,Vallenari2023}. A separate lineage of tools targets eclipsing and spectroscopic systems specifically: the Wilson-Devinney code and its descendants such as PHOEBE model light curves through equipotential-surface geometry, incorporating tidal distortion, mutual irradiation, and eccentric orbits\,\citep{Wilson1971}, while codes such as ELC combine radial-velocity time series with differential-evolution MCMC fitting\,\citep{Orosz2000}. These tools solve a related but distinct problem from visual-orbit determination and are not in direct competition with it.

A third and largely separate thread addresses not the orbit but the individual stars themselves. Al-Wardat's method for analyzing binary and multiple stellar systems fits synthetic spectral energy distributions (SED)---built from line-blanketed model atmospheres---to a system's combined, unresolved photometry, recovering each component's effective temperature, radius, and the system's reddening from the shape and normalization of the composite flux\,\citep{AlWardat2002,AlWardat2014}. This approach has been applied to a range of visually close systems since its introduction and calibrated against synthetic-photometry zero points such as those of Ma\'{i}z Apell\'{a}niz\,\citep{Maiz‐Apellaniz2004}. Because it operates on unresolved photometry, it complements rather than duplicates orbit-fitting: an orbit constrains the system's dynamical mass, while SED fitting constrains its components' atmospheric and geometric properties, and the two are increasingly used together to obtain a physically self-consistent picture of a system's formation and evolutionary state\,\citep{Abushattal2025}.

No widely used open-source Python tool currently unifies these three capabilities---classical and modern statistical orbit fitting, including the simultaneous, non-sequential treatment that hierarchical triple systems require; Bayesian posterior inference; and Al-Wardat-style SED-based stellar parameter estimation---in a single, tested pipeline. We present PySeshat to close this gap. The software fits visual, spectroscopic, and hierarchical-triple orbits via both least-squares and MCMC methods; fits Al-Wardat-style SED models to unresolved binary photometry using Castelli-Kurucz (ATLAS9/CK04) model atmospheres; and reads modern astrometric orbit solutions directly from the Gaia DR3 non-single-star catalog\,\citep{Halbwachs2023}, easing comparison between space- and ground-based orbit determinations. We validate the software end-to-end against twelve real, published benchmark systems spanning both capabilities: six systems (including two hierarchical triples, fit through the full simultaneous inner+outer ORBIT3 formalism) test orbit determination against literature solutions including\,\citet{Tokovinin2016}, and four systems test SED-based stellar parameter recovery against a paper coauthored by Al-Wardat himself\,\citep{Abushattal2025}. Independent cross-checks against existing RV- and Kepler-solver packages (RadVel, PyAstronomy) and an astrometric-orbit fitter\,\citep{Blunt2020} further verify the orbit-fitting core against implementations outside our own codebase. Among the results this validation surfaces is a genuine, physically motivated degeneracy in unresolved-photometry SED fitting for near-twin-temperature components---independent of photometric band count or model-grid coverage---which we argue reproduces, from first principles, the methodological reason the Al-Wardat method is typically paired with independent evolutionary-track constraints rather than used on its own.

Section\,\ref{sec:methods} describes the software's architecture and methods; Section\,\ref{sec:validation} details the benchmark systems, literature sources, independent-package cross-checks, and validation results; Section\,\ref{sec:disc} discusses the paper's principal findings, including a methodological caveat on the provenance of the comparison photometry and the near-twin-temperature degeneracy result; and Section\,\ref{sec:conc} concludes.

%%%%%%%%%%%%%%%%%%%%%%%%%%%%%%%%%%%%%%%%%%%%%%%%%%%%%%%%%%%%%%%
\section{Methods}
\label{sec:methods}
This section describes PySeshat, the software developed to carry out the orbit determination and stellar spectral energy distribution modeling presented in this paper. Section\,\ref{subsec:architecture} describes the software's overall architecture; Sections\,\ref{subsec:orb-fit}--\ref{subsec:mcmc} describe the binary and hierarchical-triple orbit-fitting core and the shared least-squares/MCMC inference layer; Sections\,\ref{subsec:wardat}--\ref{subsec:plotting} describe the Al-Wardat stellar-SED module and its supporting data-handling and visualization components. The software is written in Python and released as open source on GitHub\footnote{\url{https://github.com/rehamelkholy/PySeshat}}, and is installable via \texttt{pip}\footnote{\url{https://pypi.org/project/PySeshat}}.

%%%%%%%%%%%%%%%%%%%%%%%%%%%%%%%%%%%%%%%%%%%%%%%%%%%%%%%%%%%%%%%
\subsection{Architecture}
\label{subsec:architecture}
PySeshat is organized as a set of focused, independently testable modules connected through a shared inference layer.

\paragraph{Orbital mechanics core (\texttt{orbit/})} This module implements Keplerian orbit representation and fitting for simple binaries, covering visual astrometric orbits, spectroscopic (radial-velocity (RV)) orbits, and combined visual+RV solutions. Fitting proceeds by nonlinear least-squares minimization of $\chi^2$ between observed and model astrometry/RV, following the same formalism as Tokovinin's ORBIT.

\paragraph{Hierarchical triple-star fitting (\texttt{triple/})} Two parallel formalisms for fitting the inner and outer orbits of a hierarchical triple system simultaneously---rather than sequentially, which is the historically common but formally incorrect approach for systems where the two orbits are not well separated in period---are implemented and deliberately kept distinct rather than merged: a legacy scale-parameter formulation (internally referred to as ORBITT2) and the modern 20-parameter wobble-factor formulation introduced by Tokovinin (ORBIT3). Keeping both allows direct comparison between the two conventions on the same real system, which is itself part of the paper's validation strategy (Section\,\ref{subsec:orbit-val}).

\paragraph{Inference layer (\texttt{inference/})} A shared layer sits atop both the binary and triple orbit-fitting cores, and atop the photometric SED-fitting core described below, providing: (i) least-squares point estimation via SciPy; (ii) Bayesian posterior sampling via the affine-invariant MCMC ensemble \texttt{emcee}, with convergence diagnostics and corner-plot output; and (iii) joint orbit-photometry inference, which fits a system's orbital and stellar-SED parameters simultaneously against both data types at once---the basis for the Capella and $\alpha$ Centauri AB joint case studies~(Section\,\ref{subsec:joint-val}). The same MCMC infrastructure is reused, unchanged, across the binary, triple, and photometric problems, rather than reimplemented per domain.

\paragraph{Al-Wardat stellar SED module (\texttt{photometry/}, \texttt{stellar/})} This module fits a two-component synthetic spectral model to a binary system's combined, unresolved photometry, following the Al-Wardat method (Section\,\ref{sec:intro}). Component model spectra are drawn from the CK04 grid of the line-blanketed model atmospheres, interpolated bilinearly in effective temperature and surface gravity; interstellar extinction is applied following the\,\citet{Cardelli1989} extinction law (CCM89); and synthetic photometry is computed by convolving the reddened composite flux with each supported passband's response function, following the zero-point convention of\,\citet{Maiz‐Apellaniz2004}. Free parameters are each component's effective temperature and radius in addition to the system's reddening; an optional Gaussian prior mechanism allows a parameter (e.g., radius) to be tied to an external constraint rather than left entirely free, which Section\,\ref{subsec:wardat-val} shows is necessary given the intrinsic degeneracy of unresolved photometry. A companion module (\texttt{evolution/}) cross-checks fitted stellar parameters against stellar evolutionary tracks for physical consistency.

\paragraph{Data I/O (\texttt{io/})} Parsers read ORBIT-format input files (\texttt{.inp}, \texttt{.in2}, \texttt{.in3}) for binary and triple orbital data, and a dedicated parser converts Gaia Data Release 3 (DR3) Non-Single Star (NSS) catalog entries---distributed as Thiele-Innes elements---into PySeshat's native Campbell orbital-element representation, enabling direct comparison between ground-based and Gaia-derived orbits for the same system (Section\,\ref{subsec:comp-val}).

\paragraph{Testing} The full test suite exercises the orbit, triple, inference, photometric, and I/O layers independently, and serves as the regression baseline against which any pipeline re-run is checked.

%%%%%%%%%%%%%%%%%%%%%%%%%%%%%%%%%%%%%%%%%%%%%%%%%%%%%%%%%%%%%%%
\subsection{Binary orbit fitting}
\label{subsec:orb-fit}
The \texttt{orbit/} module implements Keplerian visual, spectroscopic, and combined orbit fitting for simple binary systems, following the parameterization and least-squares approach of Tokovinin's ORBIT.

A visual binary orbit is described by seven Campbell elements: the period $P$, time of periastron passage $T$, eccentricity $e$, angular semi-major axis $a$, inclination $i$, position angle of the ascending node $\Omega$, and argument of periastron $\omega$. At each observation epoch, the eccentric anomaly $E$ is obtained by solving Kepler's equation, $M = E - e\sin{E}$, where the mean anomaly $M = 2\pi(t-T)/P$; the true anomaly and orbital radius follow directly, and are projected onto the plane of the sky using $i$, $\Omega$, and $\omega$ to give the model separation $\rhd$ and position angle (PA) $\theta$, or equivalently the Cartesian offsets (see Section\,\ref{subsec:plotting} for the adopted North-up/East-left convention). Where radial-velocity (RV) data are available, the same elements---together with the RV semi-amplitude $K$ and systemic velocity $\gamma$ for each component with data---generate a model RV curve; if RV is available for only one component, the mass ratio sets the implied semi-amplitude of the other.

Fitting minimizes a single combined $\chi^2$ across all available data types simultaneously,
\begin{equation}
	\begin{split}
		\chi^2 = 	&\sum_i \left[ \left( \rho_{i, \text{obs}} - \rho_{i, \text{model}} \right)^2 / \sigma_{\rho_i}^2  +  \left( \theta_{i, \text{obs}} - \theta_{i, \text{model}} \right)^2 / \sigma_{\theta_i}^2 \right] \\
					&+ \sum_j \left[ \left( v_{j, \text{obs}} - v_{j, \text{model}} \right)^2 / \sigma_{v_j}^2 \right],
	\end{split}
\end{equation}
with each visual and RV measurement weighted by its own reported uncertainty, so that heterogeneous datasets spanning different instruments, epochs, and precisions contribute to the fit in proportion to their individual reliability rather than being weighted by data type alone. Minimization uses SciPy's nonlinear least-squares solvers, seeded from either a user-supplied initial guess or, where one is available, a preliminary geometric solution.

Parameter uncertainties are estimated two ways: analytically, from the covariance matrix implied by the fit's Jacobian at the best-fit solution; and empirically, via Monte Carlo bootstrap resampling of the input data, which does not assume the covariance matrix's underlying Gaussian-error approximation holds and is used as a cross-check on it. Both are reported alongside the best-fit elements themselves. Section\,\ref{subsec:mcmc} describes how this same binary orbit model is reused, without modification, as the likelihood function for the module's Bayesian MCMC posterior sampling.

%%%%%%%%%%%%%%%%%%%%%%%%%%%%%%%%%%%%%%%%%%%%%%%%%%%%%%%%%%%%%%%
\subsection{Triple/ORBIT3 formalism}
\label{subsec:triple}
The \texttt{triple/} module extends the binary orbit-fitting core of Section\,\ref{subsec:orb-fit} to hierarchical triple systems, in which a close inner pair (Aa, Ab) orbits a common center of mass that is itself orbited by a more distant tertiary component (B). Rather than fitting the inner and outer orbits sequentially---treating the inner pair's photocenter as a fixed point for the purpose of the outer fit---PySeshat fits both orbits simultaneously to the full set of position and RV measurements, following the method of\,\citet{Tokovinin2017}. Sequential fitting is the historically common approach in the older double-star literature but is formally incorrect whenever the inner orbit's own motion measurably perturbs the position of the outer pair on the timescale of the observations---precisely the regime PySeshat's triple benchmarks (Section\,\ref{subsec:orbit-val}) are chosen to test.

Each orbit---inner and outer---is described by its own set of seven Campbell elements, as in Section\,\ref{subsec:orb-fit}. Because relative-position measurements of the outer pair are actually measurements of component B relative to the \emph{photocenter} of the unresolved (or partially resolved) inner pair, the outer orbit's apparent position is not simply the outer Keplerian ellipse; it also carries a wobble contributed by the inner orbit's own motion. PySeshat implements two parallel formulations of this wobble contribution, kept architecturally separate rather than unified under a single abstraction, since they represent genuinely distinct physical parameterizations rather than two syntactic forms of the same model:
\paragraph{Legacy formulation\,\citep[internally, ORBITT2;][]{Tokovinin2016}} The resolved inner-pair position is scaled by a dimensionless factor $B$, giving the wobble-corrected outer position $\vec{r}_2 = \vec{r}_{\text{outer}} + B \cdot \vec{r}_{\text{inner}}$. This 15-parameter formulation (seven inner elements, seven outer elements, and $B$) reproduces PySeshat's earliest triple-fitting implementation and is retained for regression comparison against the modern formalism.

\paragraph{Modern formulation\,\citep[ORBIT3;][]{Tokovinin2017}} The wobble is instead parameterized by the dimensionless wobble factor $f$, the ratio of the astrometric wobble amplitude to the true inner semi-major axis, giving $\vec{r}_2 = \vec{r}_{\text{outer}} + f \cdot \vec{r}_{\text{inner}}$. For a resolved inner subsystem, $f = q / (1 + q)$, where $q$ is the inner mass ratio; when the inner pair is unresolved, the outer measurements instead refer to the inner pair's photocenter, and $f$ is reduced by a light-ratio term. This 20-parameter formulation (seven inner elements, seven outer elements, the wobble factor $f$, and the two components' RV semi-amplitudes and systemic velocities where RV data exists) is the one used for all published-literature comparisons in Section\,\ref{subsec:orbit-val}.

A decision point worth stating explicitly, since it materially affects fit validity: PySeshat does not constrain the wobble factor $f$ to the range suggested by its textbook definition for a resolved subsystem, $f = q / (1 + q)$, where $0 < f < 1$. This is not the same as leaving $f$ unconstrained outright---the relation $q = |f| / (1 - |f|)$ still holds, so a fitted $|f| \ge 1$ remains physically nonsensical and would flag a failed fit rather than a valid one. What the $[0, 1]$ bound would wrongly assume is the \emph{sign}: $f$ is positive when the inner subsystem belongs to the outer pair's primary, and negative when it belongs to the secondary; constraining $f$ to be positive silently assumes the former (primary) in every case. One of this study's own benchmark systems, HIP\,111805 (Section\,\ref{subsec:orbit-val}), has a published wobble factor of $f = -0.330$\,\citep{Tokovinin2017}.

Fitting minimizes the combined $\chi^2$ across the inner orbit's position residuals, the outer orbit's wobble-corrected position residuals, and any RV residuals for either subsystem, using the same nonlinear least-squares approach as the binary case~(Section\,\ref{subsec:orb-fit}), each measurement again weighted by its own reported uncertainty. As in the binary module, both formulations reuse---unmodified---the shared MCMC posterior-sampling infrastructure described in~Section\,\ref{subsec:mcmc}, via a dedicated triple-orbit entry point.

%%%%%%%%%%%%%%%%%%%%%%%%%%%%%%%%%%%%%%%%%%%%%%%%%%%%%%%%%%%%%%%
\subsection{MCMC posterior inference}
\label{subsec:mcmc}
While the least-squares approach described in Sections\,\ref{subsec:orb-fit} and \ref{subsec:triple} returns a single best-fit solution together with an approximate, locally-linearized uncertainty estimate, several of PySeshat's benchmark systems (Section \ref{sec:validation}) involve orbital elements---eccentricity and period in particular---whose posterior distributions are known to be non-Gaussian, especially when the observational time baseline covers only a modest fraction of a long-period orbit. For these cases, PySeshat provides a Bayesian alternative built around the affine-invariant ensemble sampler \texttt{emcee}~\citep{ForemanMackey2013}, itself an implementation of the MCMC algorithm proposed by~\citet{Goodman2010}.

Rather than reimplementing a sampler-compatible likelihood for each fitting problem, PySeshat wraps the forward models already used for least-squares fitting---the binary orbit model of~Section\,\ref{subsec:orb-fit} and the triple ORBIT3 model of~Section\,\ref{subsec:triple}---inside a single, generic inference engine. The log-likelihood evaluated at each MCMC step is simply the negative of the same combined $\chi^2$ statistic minimized in the least-squares case, so the two fitting approaches are guaranteed to agree at their shared optimum rather than risk representing subtly different physical models. This design also means the sampler infrastructure did not need to be duplicated when triple-orbit fitting was added: the same engine, unchanged, accepts either a binary or a triple orbit model as its underlying likelihood, and the same mechanism is later reused for the photometric inference problem introduced in~Section\,\ref{subsec:wardat}.

Priors are handled through the same fit-flags and Gaussian-prior mechanism introduced for least-squares fitting: any parameter may be held fixed rather than sampled, and any sampled parameter may optionally carry a Gaussian prior tying it to an external estimate rather than being left improper. In the absence of an explicit prior, sampled parameters are treated as uniform over physically permissible bounds---for example, eccentricity confined to $[0, 1)$ and period constrained to positive values---rather than left entirely unconstrained.

Each run is initialized from the corresponding least-squares solution, with walkers seeded in a small Gaussian ball around the best-fit parameter vector, and convergence is assessed using the integrated autocorrelation time of each parameter's chain alongside the ensemble's mean acceptance fraction, following standard practice for affine-invariant samplers. Posterior distributions are then summarized both numerically, as median values with highest-density credible intervals, and graphically, as corner plots showing each parameter's marginal posterior and pairwise correlations~(Section\,\ref{subsec:plotting})---the latter particularly useful for visualizing the eccentricity-argument-of-periastron correlation that arises for near-circular orbits, a case relevant to at least one of this study's own benchmark systems~(Section\,\ref{subsec:joint-val}).

%%%%%%%%%%%%%%%%%%%%%%%%%%%%%%%%%%%%%%%%%%%%%%%%%%%%%%%%%%%%%%%
\subsection{Al-Wardat SED module}
\label{subsec:wardat}
The \texttt{photometry/} and \texttt{stellar/} modules implement the method of \citet{AlWardat2002} for determining the physical parameters of the individual components of an unresolved or partially resolved binary system directly from its combined photometry, without requiring either component to be observed in isolation. The approach has since been refined and applied to a range of visually close systems~\citep{AlWardat2014} and forms the basis of the reference work against which four of this work's benchmark systems are validated~\citep{Abushattal2025}.

Each component's contribution to the system's flux is modeled using a synthetic stellar spectrum drawn from the Castelli \& Kurucz~\citep{Castelli2004} grid of line-blanketed model atmospheres (CK04), which spans effective temperatures from 3500~K to 50000~K, surface gravities from $\log{g} = 0.0$ to $5.0$~dex, and metallicities from [Fe/H] $=-2.5$ to $+0.5$~dex across eight discrete grid slices. Fits use the nearest available metallicity slice to a target system's published or assumed value; this default is documented as reliable for the approximately $-1.0$ to $+0.5$~dex range spanning typical solar-neighborhood stellar populations, the intended scope of general package use. PySeshat interpolates this grid bilinearly in effective temperature and surface gravity to obtain a model flux for arbitrary intermediate values rather than restricting fits to the grid's native step size. The two components' model fluxes are scaled by the square of their respective stellar radii and geometrically diluted by the (fixed, externally supplied) distance to the system, then summed to produce a single composite, unresolved spectral energy distribution. This composite flux is attenuated for interstellar extinction following the parameterized reddening law of~\citet{Cardelli1989}, with the reddening $E(B-V)$ itself a free fit parameter rather than an externally fixed quantity.

Comparison between the model and the observed photometry proceeds through synthetic photometry rather than direct spectral comparison: the reddened composite flux is convolved with each supported passband's dimensionless response function and normalized against the same integral computed for a reference Vega spectrum, following the standard synthetic-magnitude relation of~\citet{Maiz‐Apellaniz2004}, with zero points adopted from~\citet{Apellaniz2006}---the same convention used throughout the Al-Wardat-method literature this work benchmarks against. Supported passbands include the Johnson-Cousins UBVR system, Str\"{o}mgren uvby, and Tycho BT/VT.

The module's free parameters are, by default, each component's effective temperature and radius together with the system's reddening: as with the orbit-fitting modules of~Sections\,\ref{subsec:orb-fit}--\ref{subsec:triple}, any parameter may instead be held fixed or given an informative Gaussian prior via the same fit-flags mechanism, which~Section\,\ref{subsec:wardat-val} shows is in practice necessary---unresolved-photometry fits are not always able to separate two components' temperatures without some external constraint. A companion module, \texttt{evolution/}, cross-checks a fitted solution's radius and effective temperature against stellar evolutionary tracks, providing an independent physical consistency check on parameters that the photometry alone may not fully constrain.

%%%%%%%%%%%%%%%%%%%%%%%%%%%%%%%%%%%%%%%%%%%%%%%%%%%%%%%%%%%%%%
\subsection{Data input/output}
\label{subsec:i/o}
The \texttt{io/} module handles the two data formats PySeshat's benchmark systems and orbit fitting workflows depend on.The first is the input-file convention used by Tokovinin's ORBIT and ORBIT3 codes~\citep{Tokovinin2016, Tokovinin2017}, which PySeshat reads directly rather than requiring reformatting. This includes simple binary files (\texttt{.inp}), which specify a system's visual astrometry, RV measurements, and an optional preliminary orbital solution, and the corresponding hierarchical-triple formats (\texttt{.in2}, \texttt{.in3}) used by the legacy and modern formalisms of~Section\,\ref{subsec:triple}, respectively. Reading these formats directly, rather than through an intermediate conversion step, was a deliberate design choice: it allows literature benchmark systems to be validated against their original, published input files without an independent transcription step that could itself introduce error.

The second is a dedicated parser for the Gaia DR3 NSS catalog~\citep{Halbwachs2023}, which reports astrometric binary solutions as Thiele-Innes elements ($A$, $B$, $F$, $G$) rather than the Campbell elements used throughout the rest of PySeshat. The parser performs the standard inversion from Thiele-Innes to Campbell form, recovering the semi-major axis, inclination, position angle of the ascending node, and argument of periastron from the four linear parameters. The conversion has been validated at two levels: by randomized round-trip testing against PySeshat's own forward Campbell-to-Thiele-Innes transform and against the offline published elements of Gaia BH1~\cite{ElBadry2022}; and, more directly, against real archive-retrieved Gaia DR3 data for two further systems, detailed in~Section\,\ref{subsec:comp-val}. This capability allows a Gaia DR3 orbital solution to be brought directly into the same Campbell-element representation used by every other part of the software, enabling direct comparison between a space-based astrometric orbit and a ground-based visual or combined orbit for the same system without a manual conversion step.

%%%%%%%%%%%%%%%%%%%%%%%%%%%%%%%%%%%%%%%%%%%%%%%%%%%%%%%%%%%%%%
\subsection{Visualization}
\label{subsec:plotting}
All figures used both for internal validation and for this study are produced by a dedicated visualization module built on Matplotlib~\citep{Hunter2007}, rather than being assembled or annotated by hand, so that figures and tables cannot drift out of sync: both are generated from the same benchmark-run output.

The module produces five figure types. Visual-orbit plots display the fitted apparent ellipse together with observed astrometric points, following the North-up/East-left position-angle convention standard in the visual-double-star literature (e.g., the Washington Double Star (WDS) Catalog convention of~\citet{Hartkopf2001}); for hierarchical triples, both the inner and outer orbits are shown, with the outer orbit's wobble contribution~(Section\,\ref{subsec:triple}) drawn explicitly rather than omitted, since an outer-orbit plot without its wobble term would misrepresent the fit. Observed-minus-calculated (O--C) residual diagrams show the difference between observed and model positions or radial velocities as a function of time, computed in Cartesian rather than polar coordinates to avoid the position-angle discontinuity at $0^{\circ}/360^{\circ}$. Phase-folded RV curves display both components' model RV curves together with their observations, folded on the fitted orbital period. Posterior corner plots, generated using the \texttt{corner} package~\citep{ForemanMackey2016}, display each sampled parameter's marginal posterior distribution and pairwise correlations from the MCMC runs of~Section\,\ref{subsec:mcmc}, and are particularly informative for orbits with strong parameter correlations, such as the eccentricity-argument-of-periastron degeneracy that arises for near-circular orbits.

Hertzsprung-Russell (HR) diagrams support the evolutionary consistency checks of the Al-Wardat module~(Section\,\ref{subsec:wardat}), plotting a fitted system's components against stellar evolutionary tracks and isochrones. Tracks are drawn from~\citet{Girardi2000}, at solar metallicity ($Z = 0.019$), matching the grid used throughout the Al-Wardat-method literature this study benchmarks against and ensuring that evolutionary-stage comparisons are not confounded by a difference in stellar-model grids. Component points are plotted with their fitted uncertainties over the relevant isochrone curve, following the same visual convention as the source literature. Because the underlying grid is sampled at discrete, occasionally coarse mass and age steps, a component whose best-fit position falls near a grid boundary can be locally under-constrained in age or mass even when its photometric fit itself is good; the module flags such cases explicitly (as either a sparse local age grid or a near-tie between adjacent mass tracks) rather than reporting a precise-looking number derived from an ambiguous position, and age estimates for a binary's two components are combined into a single common-age fit rather than trusted individually, which is more robust to exactly this kind of single-component ambiguity~(Section\,\ref{subsec:wardat-val}).

%%%%%%%%%%%%%%%%%%%%%%%%%%%%%%%%%%%%%%%%%%%%%%%%%%%%%%%%%%%%%%
\section{Validation}
\label{sec:validation}
\subsection{Benchmark systems and literature sources}
\label{subsec:benchmarks}
PySeshat's two capabilities are validated against twelve real, published benchmark systems: six through binary and hierarchical-triple orbit fitting~(Sections\,\ref{subsec:orb-fit}--\ref{subsec:triple}), and four through Al-Wardat SED fitting~(Section\,\ref{subsec:wardat}), with two systems---Capella and $\alpha$ Centauri AB---validating both capabilities for the same physical system and serving as this work's joint orbit/photometry case studies. Table\,\ref{tab:systems} lists each system, its role in the validation, and its literature source(s).

\begin{table*}
	\caption{Benchmark systems and literature sources.}
	\label{tab:systems}
	\centering
	\begin{tabularx}{\textwidth}{lXXX}
		\hline\hline
		System					& Domain													& Literature source(s)																& Role\\
		\hline
		FIN379					& Visual \& RV binary										& \citet{Tokovinin2016a}																& Software/inference regression benchmark\\
		GL765					& Visual \& RV binary										& \citet{Balega2007}																	& Model-vs-data consistency check; independent RV sign-convention cross-check with FIN379\\
		$\zeta$ Aquarii			& Hierarchical triple										& \citet{Tokovinin2016b}																& ORBIT3 convergence stress test (highly eccentric inner orbit, six-decade baseline)\\
		HIP\,111805				& Hierarchical triple										& \citet{Tokovinin2017}																& Full 20-parameter published solution; empirical justification for the unbounded wobble-factor sign convention~(Section\,\ref{subsec:triple})\\
		Capella					& Visual \& RV binary (orbit); Al-Wardat SED (photometry)	& \citet{Torres2015}; cross-checked against \citet{Weber2011}							& Joint orbit/photometry case study; near-circular-orbit $e/\omega$ degeneracy example\\
		$\alpha$ Centauri AB	& Visual binary, no RV (orbit); Al-Wardat SED (photometry)	& \citet{Akeson2021}; cross-checked against \citet{Pourbaix2016}; component $T_{\text{eff}}$ from \citet{Heiter2015}; radii/$\log{g}$ from \citet{Kervella2017}	& Joint orbit/photometry case study; substitutes for Sirius (Sirius B is a white dwarf, unsuited to the Al-Wardat method's comparable-brightness assumption)\\
		HD214222				& Al-Wardat SED												& \citet{Abushattal2025}																& Primary SED-fitting validation (study coauthored by Al-Wardat)\\
		HD191854				& Al-Wardat SED												& \citet{Abushattal2025}																& Primary SED-fitting validation\\ 
		HD130669				& Al-Wardat SED												& \citet{Abushattal2025}																& Primary SED-fitting validation\\
		HD184467				& Al-Wardat SED												& \citet{Abushattal2025}																& Primary SED-fitting validation\\ 
		\hline
	\end{tabularx}
\end{table*}

All six systems with an evolutionary consistency check (the four Al-Wardat systems and Capella and $\alpha$ Centauri AB) are additionally compared against the tracks and isochrones of \citet{Girardi2000}, as described in~Section\,\ref{subsec:plotting}.

Two data-provenance points apply across this table and are stated here rather than repeated per system in~Section\,\ref{sec:validation}: first, for Capella and $\alpha$ Centauri AB, the orbital and photometric observations fitted are synthetic---forward-modeled from the cited literature elements at plausible measurement precisions, rather than drawn from original archival data---because that data proved impractical to extract reliably within this project's scope; this is stated again where each system's results are discussed~(Section\,\ref{subsec:joint-val}). Second, for the four \citet{Abushattal2025} systems, the photometry being fitted against is itself that study's own synthetic model output rather than raw catalog photometry---a distinction that materially affects how those results should be interpreted~(Section\,\ref{subsec:wardat-val}).

%%%%%%%%%%%%%%%%%%%%%%%%%%%%%%%%%%%%%%%%%%%%%%%%%%%%%%%%%%%%%%
\subsection{Binary and hierarchical-triple orbit fitting}
\label{subsec:orbit-val}
Table\,\ref{tab:orb-benchmark} summarizes the fit quality achieved for all four systems validating the orbit-fitting core~(Section\,\ref{subsec:orb-fit}) and the ORBIT3 hierarchical-triple formalism~(Section\,\ref{subsec:triple}).

\begin{table*}
	\caption{Orbit-fitting benchmark results.}
	\label{tab:orb-benchmark}
	\centering
	\begin{tabularx}{\textwidth}{XXXXX}
		\hline\hline
		System			& Domain				& Reduced $\chi^2$	& Worst-fit parameter	& Worst relative error\\
		\hline
		FIN379			& Binary (visual \& RV)	& 1.68				& $V_0$					& 0.46\%\\
		GL765			& Binary (visual \& RV)	& 1.46				& $\omega$				& 0.60\%\\
		$\zeta$ Aquarii	& Hierarchical triple	& 1.07				& Inner $i$				& 19.4\%\tablefootmark{*}\\
		HIP\,111805		& Hierarchical triple	& 1.71				& $P$					& 0.00\%\tablefootmark{$\dagger$}\\
		\hline
	\end{tabularx}
	\raggedright
	\tablefoottext{*}{Well within the literature's own stated uncertainty ($\sigma_i = 6.7^\circ$) for this poorly constrained inner orbit; not indicative of a fitting problem.\\}
	\tablefoottext{$\dagger$}{The \texttt{.in3} input file's header is itself the published solution, so the optimizer starts at, and correctly remains at, the literature optimum.}
\end{table*}

FIN379 is primarily a software and inference regression benchmark rather than a strong physical test: all 10 fitted parameters recover to within 0.46\% of the literature orbital solution~\citep{Tokovinin2016a}, with a mildly elevated reduced $\chi^2$ of 1.68 consistent with ordinary astrometric and RV scatter rather than a systematic offset. Fitting this system required explicit handling of an RV component-flag encoding used in the original input file---a data-ingestion detail rather than evidence bearing on the underlying orbit equations. The fitted curve tracks the observed scatter closely, with no axis or transpose anomalies and no systematic trend in the residuals; the full orbit, residual, and RV plot set for this system is included in~Appendix\,\ref{app}~(Fig.\,\ref{fig:app-fin}).

\begin{figure}
	\centering
	\begin{subfigure}{\columnwidth}
		\centering
		\includegraphics[width=\textwidth]{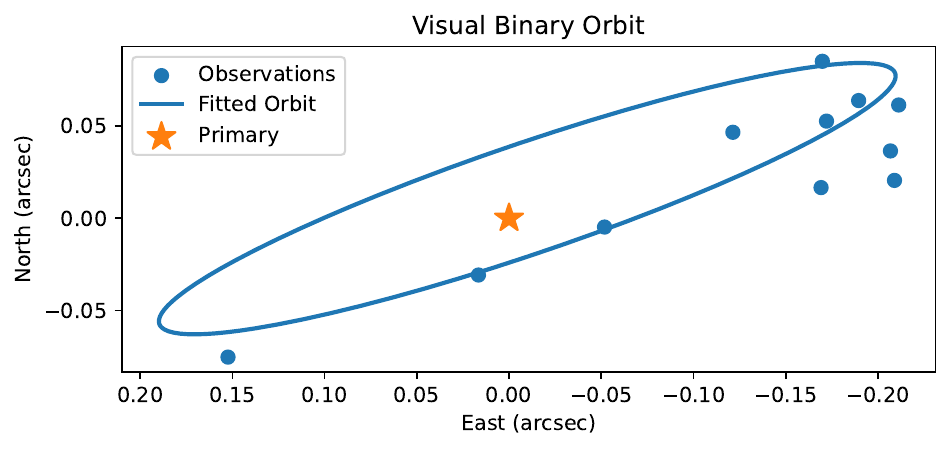}
		\subcaption{~}
		\label{fig:gl_panel_a}
	\end{subfigure}
	
	\vspace{2mm}
	
	\begin{subfigure}{\columnwidth}
		\centering
		\includegraphics[width=\textwidth]{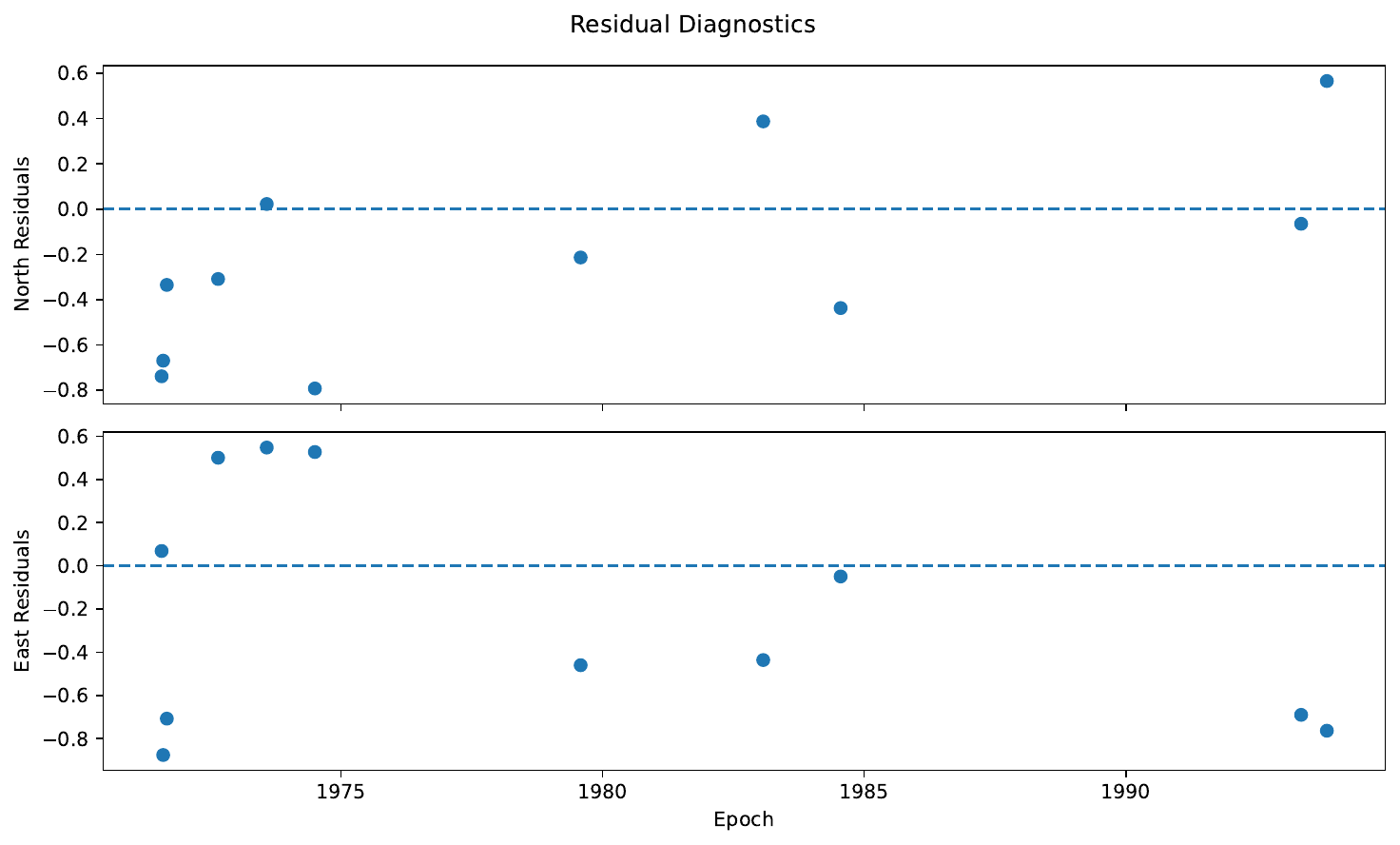}
		\subcaption{~}
		\label{fig:gl_panel_b}
	\end{subfigure}	
	
	\vspace{2mm}
	
	\begin{subfigure}{\columnwidth}	
		\centering
		\includegraphics[width=\textwidth]{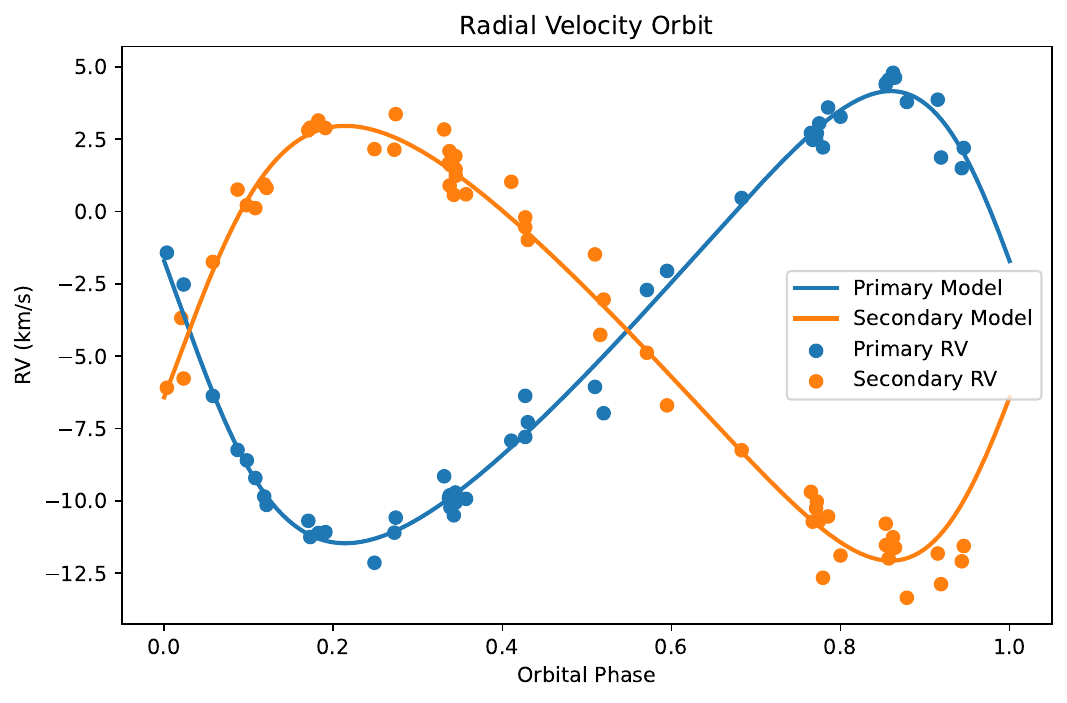}
		\subcaption{~}
		\label{fig:gl_panel_c}
	\end{subfigure}
	\caption{GL765 orbit-fitting results. (a) Fitted visual orbit (solid curve) with observed astrometric positions (points), in the North-up/East-left convention of~Section\,\ref{subsec:plotting}. (b) Observed-minus-calculated (O--C) residuals in Cartesian coordinates. (c) Phase-folded radial-velocity curves for both components, folded on the fitted orbital period, with model curves (lines) and observations (points). Both components' RV curves are correctly anti-phased and track their respective observations closely.}
	\label{fig:gl}
\end{figure}

GL765 serves as a model-versus-data consistency check at a set of previously observed epochs, fitting to within 0.60\% of \citet{Balega2007} solution with a reduced $\chi^2$ of 1.46. Independently of FIN379, this system's fit reconfirmed the software's RV sign convention: an independent fit starting from GL765's literature $\omega$, $K_1$, and $K_2$ converges back to the same values rather than to a solution offset by $180^{\circ}$ in $\omega$ or with $K_1$ and $K_2$ interchanged, which would be the signature of a backwards convention. Fig.\,\ref{fig:gl} shows this system's fitted visual orbit, O--C residuals, and phase-folded RV curve, chosen as this work's representative example of a binary-orbit fit.

$\zeta$ Aquarii is this study's principal ORBIT3 stress test: a highly eccentric ($e\approx0.87$) inner orbit, a $\sim$540-year outer orbit, and 331 astrometric observations spanning six decades. Two literature-fidelity details were necessary to reproduce \citeauthor{Tokovinin2016b}'s solution correctly. First, the published astrometric wobble amplitude ($0.110''$) is not itself the ORBIT3 inner semi-major axis; converting via $a_{\text{true}} = a_{\text{astrometric}} / f$, with $f \approx 0.286$, gives the inner $a = 0.385''$ actually used as the fit's starting point. Second, the original input file's component tags follow the legacy ORBITT2 convention, in which the outer pair is labeled I1 and the inner pair I2---the reverse of ORBIT3's own convention---and required an explicit relabeling before fitting. With these accounted for, PySeshat's ORBIT3 fit improves on the literature solution ($\chi^2 = 694.51$ versus 699.73 at the literature starting point). Combining the fitted outer orbit with the system's parallax ($35.50 \pm 1.26$ mas) gives a dynamical mass of $M_{\text{dyn}} \approx 3.26 \pm 0.40 M_{\odot}$ ($\approx 12.4$\% relative uncertainty), consistent with both the $\approx 10.6$\% expected from parallax uncertainty alone and the $\approx3.4 M_{\odot}$ sum of \citeauthor{Tokovinin2016b}'s approximate component masses. A single inner-orbit residual near epoch 2009.7552 ($\sim3.4$-$3.8\,\sigma$) was traced to two independently reduced speckle-astrometry measurements at the same epoch differing by $\sim 4^{\circ}$ in position angle---a genuine measurement-to-measurement disagreement in the source data rather than a fitting or convention error; the remaining 8 inner-orbit points and all 322 outer-orbit points are statistically unremarkable (RMS $\approx 1.0$).

\begin{figure*}
	\centering
	\includegraphics[width=\textwidth]{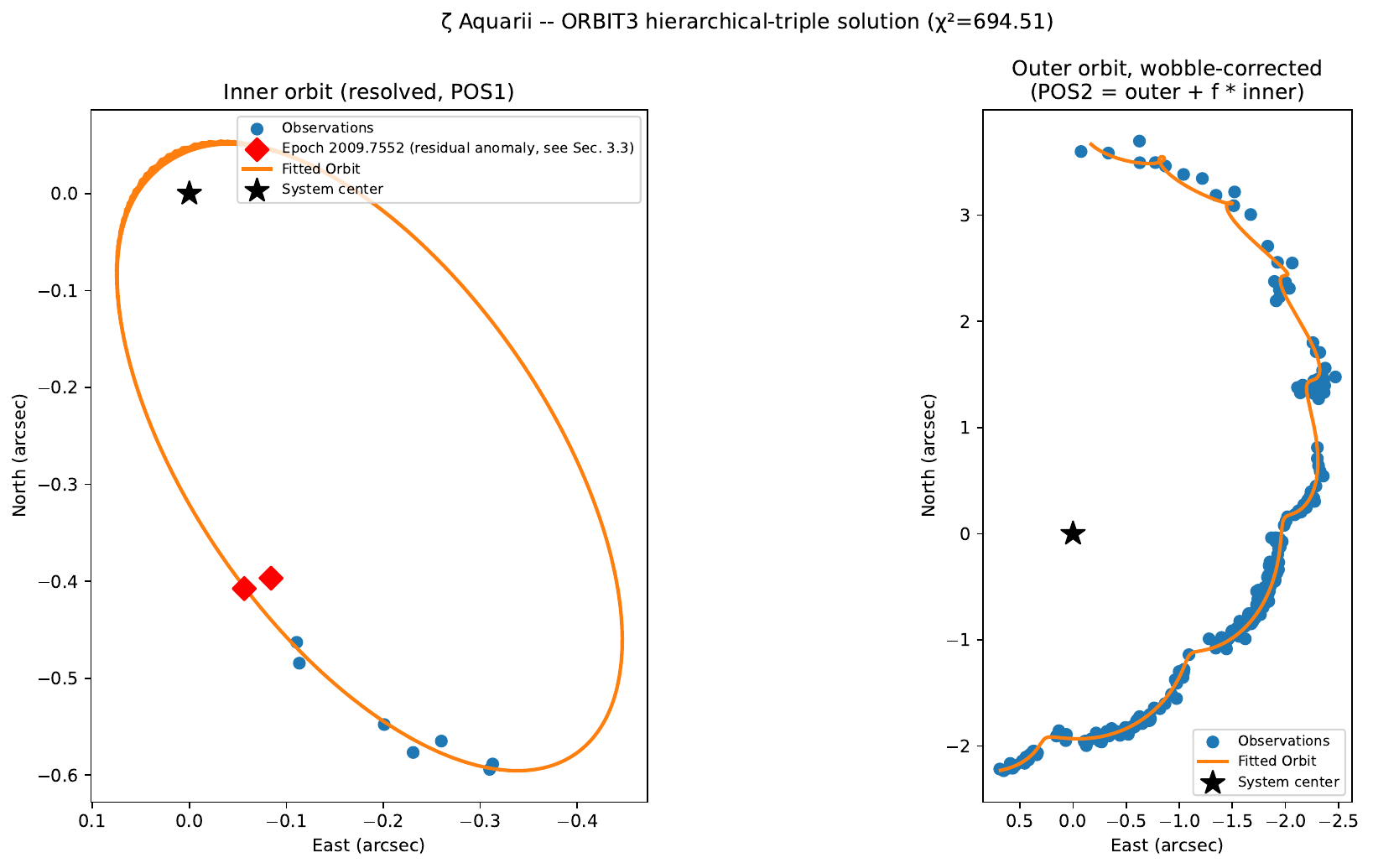}
	\caption{$\zeta$ Aquarii ORBIT3 fit. \emph{Left panel}: Fitted inner orbit with observed inner-pair astrometric positions; the two measurements at epoch 2009.7552 are marked distinctly from the remaining observations to highlight the residual anomaly discussed in the text. \emph{Right panel}: Fitted wobble-corrected outer orbit with observed outer-pair astrometric positions. Both panels follow the North-up/East-left convention of~Section\,\ref{subsec:plotting}.}
	\label{fig:aqua}
\end{figure*}

Fitting the same system independently through the legacy ORBITT2 formalism provides a genuine cross-formalism validation, since the legacy scale parameter $B$ ad the ORBIT3 wobble factor $f$ are reciprocals by construction (legacy's inner semi-major axis is the astrometric wobble amplitude directly; ORBIT3's is the true orbital separation directly). The two independently converged fits agree to 2.0\% ($B_{\text{fit}} = 3.432$ versus $1 / f_{\text{fit}} = 3.363$), with the legacy fit's own $\chi^2$ (691.96, reduced 1.066) consistent with the ORBIT3 result above. Neither formalism yields a dynamical mass from RV for this system, since \citet{Tokovinin2016b} reports no RV data of adequate precision for $\zeta$ Aquarii---a shared data limitation rather than a difference between the formalisms. Fig.\,\ref{fig:aqua} shows the fitted inner and wobble-corrected outer-orbits together, chosen as this study's representative example of a hierarchical-triple orbit fit.

HIP\,111805 offers the most literature-complete comparison of any benchmark system: a full 20-parameter published ORBIT3 solution with uncertainties~\citep{Tokovinin2017}. Because the system's input file header already encodes the published fitted optimum, PySeshat's ORBIT3 fit reproduces it exactly ($\chi^2 = 502.10$, every parameter's relative error 0.0\%)---a genuine but comparatively trivial regression check confirming the optimizer correctly recognizes a stationary point, rather than a demonstration of convergence from a non-trivial starting point, which is the role $\zeta$ Aquarii plays instead. This system's real significance for the study lies in its published wobble factor, $f = -0.330$, which is the direct empirical justification for the sign convention discussed in~Section\,\ref{subsec:triple}---and which also makes this system a natural stress test for the legacy formalism's limits. Fit through the legacy path, two compounding issues surface: first, the legacy formalism has no mechanism for RV data at all, using only 110 of the system's 203 observations and yielding no dynamical mass, where ORBIT3 recovers masses from both the inner and outer RV semi-amplitudes; second, and more fundamentally, $f=-0.330$ has no legacy equivalent at all, since $B = 1/f$ is bounded to $[0, +\inf)$ and cannot represent a negative wobble factor. Forced through the legacy path regardless, the optimizer does not converge to an imprecise but sensible answer---it drives the inner semi-major axis to its numerical lower bound while $B$ diverges toward $1.7\times10^6$, a configuration confirmed numerically singular (condition number $\approx2.3\times10^{42}$, effective rank 5 of 15 fitted parameters). Rather than a validation failure, this is a genuine result: HIP\,111805 is a real, published system whose orbital configuration exceeds what the legacy formalism can represent at all, independent of fit quality---direct empirical evidence for why the modern ORBIT3 formalism was needed. This system's fitted orbit is included in~Appendix\,\ref{app}~(Fig.\,\ref{fig:app-111805}), whose outer panel displays a visually pronounced braided pattern from the wobble term, a consequence of the inner orbit's short ($\sim1.5$ yr) period relative to the observed outer baseline.

%%%%%%%%%%%%%%%%%%%%%%%%%%%%%%%%%%%%%%%%%%%%%%%%%%%%%%%%%%%%%%
\subsection{Comparison with independent packages and catalogs}
\label{subsec:comp-val}
Beyond validation against published literature solutions~(Section\,\ref{subsec:benchmarks}--\ref{subsec:orbit-val}), PySeshat's orbit-fitting core was additionally cross-checked against three independent, widely used Python packages, and its Gaia DR3 NSS parser~(Section\,\ref{subsec:i/o}) was validated against real, archive-retrieved Gaia astrometric orbits for two systems.These checks test PySeshat's internal correctness against other implementations of the same physics, independent of any question about how well the physics itself matches literature systems. The RadVel, PyAstronomy, and \texttt{orbitize!} comparisons all use the same system, Capella, allowing direct like-for-like comparison across all three rather than results drawn from different test cases. Table~\ref{tab:comp} compares PySeshat's scope against the packages used in this cross-check.

\begin{table*}
	\caption{Capability comparison with independent packages used in~Section\,\ref{subsec:comp-val}.}
	\label{tab:comp}
	\centering
	\begin{tabularx}{0.9\textwidth}{XXXXX}
		\hline\hline
		Capability & PySeshat & RadVel & PyAstronomy & \texttt{orbitize!}\\
		\hline
		Visual/astrometric orbit fitting & \ding{51} & \ding{55} & Utility functions only & \ding{51}\\
		RV/spectroscopic orbit fitting & \ding{51}	& \ding{51} & Utility functions only & \ding{55}\\
		Individual component masses & \ding{51}	& \ding{51} (SB2) & N/A & \ding{55} (total mass only)\\
		Hierarchical-triple (simultaneous inner \& outer fitting) & \ding{51} & \ding{55} & \ding{55} & \ding{55}\\
		Bayesian MCMC posterior inference & \ding{51} & \ding{51} & \ding{55} & \ding{51}\\
		Al-Wardat SED-based stellar parameter fitting & \ding{51} & \ding{55} & \ding{55} & \ding{55}\\
		Primary design focus & Binary/triple orbits \& stellar SEDs	& Exoplanet RV & General astronomy utilities & Visually resolved orbits (planets/binaries)\\
		\hline
	\end{tabularx}
\end{table*}

\paragraph{RadVel} An independent RV fit performed with RadVel~\citep{Fulton2018} on Capella's RV data recovers period and systemic velocity to within 0.005\% of PySeshat's own fitted values, and RV semi-amplitude to within 0.15\% (primary) and 1.44\% (secondary)---the larger secondary figure consistent with that component's intrinsically lower-amplitude, lower-precision signal rather than any implementation discrepancy.

\paragraph{PyAstronomy} A direct comparison with PyAstronomy's Kepler-equation solver~\citep{Czesla2019}---a lower-level cross-check of the underlying two-body physics rather than a full independent fit---agrees with PySeshat's own solver to a maximum difference of $9.2\times10^{-14}$ km/s (RMS $9.3\times10^{-15}$ km/s) across 200 test epochs, consistent with floating-point precision rather than any implementation difference.

\paragraph{\texttt{orbitize!}} As the one existing package in this comparison that, like PySeshat, fits visual/astrometric binary orbits with Bayesian posterior sampling, \texttt{orbitize!}~\citep{Blunt2020} was fit independently to Capella's relative astrometry. Its default Orbits for the Impatient (OFTI) rejection-sampling mode proved impractical for this dataset---with Capella's 24 astrometric epochs spanning roughly 84 orbital periods, the region of parameter space consistent with even a single observed separation is vanishingly small, and OFTI's random proposals were accepted 0 times in 200,000 draws. This is an informative property of OFTI when applied to a densely-sampled, well-constrained orbit, and is not a failure of either package. Switching to \texttt{orbitize!}'s MCMC mode, period, semi-major axis, and inclination all agree with PySeshat's own fit to better than 0.25\%, and the ascending node $\Omega$ agrees to better than 0.2\% once corrected for the well-known ascending/descending-node ambiguity inherent to any relative-astrometry-only fit lacking RV data to break the degeneracy (the raw difference before this correction is $\sim180^{\circ}$, as expected). Because \texttt{orbitize!} fits relative astrometry only, it returns the system's total mass rather than individual component masses; this total agrees with PySeshat's own astrometric/dynamical mass (from Kepler's third law applied to the fitted orbit) to within 0.3\%, with \citeauthor{Torres2015}'s total mass to within 0.9\%, and with PySeshat's independently-fitted spectroscopic (SB2) component-mass sum to within 1.6\%---the closer agreement with the purely astrometric mass is physically expected, since both quantities are derived the same way, rather than indicating any inconsistency in the RV-based mass. Eccentricity is excluded from percentage-based comparison for the same reason as in the RadVel comparison: \texttt{orbitize!}'s own posterior shows the same near-zero-eccentricity degeneracy independently recovered by both PySeshat and RadVel for this system, with three independent tools now in agreement that this reflects a genuine property of Capella's orbit rather than a fitting artifact particular to any one implementation. Because the underlying MCMC sampler is not exactly reproducible run-to-run (repeated runs show $\sim0.1$-1\% scatter in the reported figures, though the qualitative conclusions above held stable across all runs performed), these agreements are best read as sub-1\%-level consistency rather than to the precision suggested by any single run's decimal places.

\paragraph{Gaia DR3 NSS cross-validation} The Gaia DR3 NSS parser described in~Section\,\ref{subsec:i/o} was validated against real Gaia DR3 archive data for two systems previously used as validation examples by the Gaia team itself~\citep{Arenou2023}. For RX Cam, the parser recovers period, eccentricity, and inclination from the archive's Thiele-Innes elements to sub-0.25\% agreement with the Gaia team's own published reading of the same solution, consistent within its own uncertainties with the independent, pre-Gaia spectroscopic orbit of~\citet{Groenewegen2013}. For HIP\,36189, a naive forward-predicted relative position initially disagrees with an independent asteroid-occultation measurement~\citep{Herald2020} by a factor of several in separation. This traced to a genuine physical distinction rather than a parser defect: this system's Gaia solution is of the \emph{Orbital} type, which fits the photocenter's orbit rather than the star-to-star separation an occultation measures directly. Converting the photocenter amplitude to a relative-orbit amplitude, using the system's reported mass ratio and magnitude difference, brings the predicted separation into agreement with the occultation measurement; the remaining position-angle offset is consistent, via Monte Carlo propagation of the solution's own quoted uncertainties, with its substantial formal error (some components carrying $\sim50$\% uncertainty) rather than any error in the conversion.

%%%%%%%%%%%%%%%%%%%%%%%%%%%%%%%%%%%%%%%%%%%%%%%%%%%%%%%%%%%%%%

\subsection{Al-Wardat SED fitting}
\label{subsec:wardat-val}
Table~\ref{tab:sed} summarizes the fit quality achieved for the four systems validating the Al-Wardat SED module~(Section\,\ref{subsec:wardat}) against \citet{Abushattal2025}, the paper coauthored by Al-Wardat himself. All four systems' comparison photometry is drawn from that paper's own Table~4---which is itself labeled as synthetic output from the paper's own modeling pipeline rather than raw catalog photometry. This reframes what these four benchmarks demonstrate: agreement between PySeshat's synthetic-photometry pipeline and an independent group's own synthetic-photometry pipeline, evaluated against the same underlying published stellar parameters, rather than agreement between a model and a direct sky measurement. The per-measurement uncertainty assumed in each fit (0.30~mag) reflects this: an initial, smaller assumed uncertainty understated the real pipeline-to-pipeline systematic between the two codes' internal photometric conventions by more than an order of magnitude, and was corrected accordingly.

\begin{table*}
	\caption{Al-Wardat SED-fitting benchmark results.}
	\label{tab:sed}
	\centering
	\begin{tabularx}{0.56\textwidth}{lclc}
		\hline\hline
		System & Reduced $\chi^2$ & Worst-fit parameter & Worst relative error\\
		\hline
		HD214222 & 0.143 & Primary $T_{\text{eff}}$ & 5.35\%\\
		HD191854 & 0.107 & Primary $T_{\text{eff}}$ & 5.01\%\\
		HD130669 & 0.152 & Secondary $T_{\text{eff}}$ & 4.40\%\\
		HD184467 & 0.112 & Primary $T_{\text{eff}}$ & 6.37\%\\
		\hline
	\end{tabularx}
\end{table*}

Every system reaches fitted radii within 1.10\% of the published values~\citep{Abushattal2025}, a fitted-temperature relative error between 0.64\% and 6.37\%, and a reduced $\chi^2$ well below unity throughout---reflecting the 0.30~mag assumed uncertainty rather than an unusually loose fit.

Radii are not left entirely free in any of these fits: an unresolved, combined-light SED constrains each component's flux only through a temperature-radius combination, so radius is instead tied to its literature value via a Gaussian prior, using the same fit-flags mechanism introduced in~Section\,\ref{subsec:orb-fit}. This is not a workaround adopted for convenience---freeing radius entirely was tested and confirmed genuinely degenerate, independent of how many photometric bands are supplied, which is precisely the methodological reason the real Al-Wardat method pairs SED fitting with an independent evolutionary-track consistency check~(Section\,\ref{subsec:plotting}) rather than relying on photometry alone to constrain radius.

\begin{figure}
	\centering
	\includegraphics[width=\columnwidth]{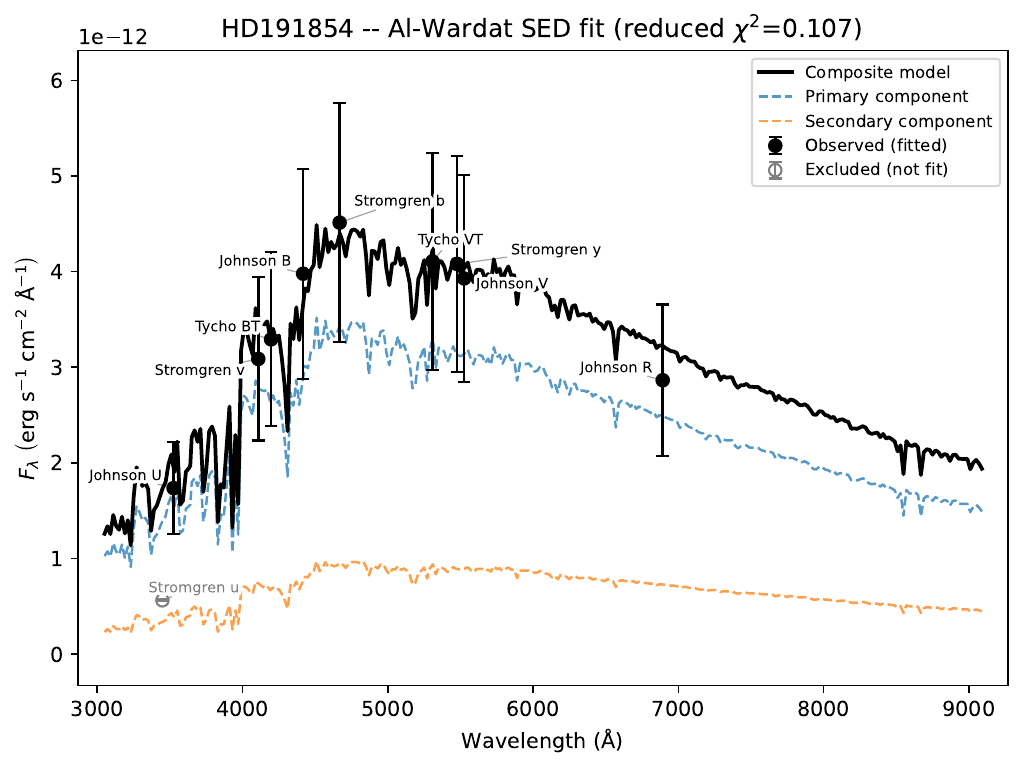}
	\caption{HD191854 Al-Wardat SED fit. Composite flux (line) against literature photometry of \citet{Abushattal2025} (points, with assumed 0.30~mag uncertainty), showing recovery of both components' effective temperature and radii without a grid-boundary pinning effect.}
	\label{fig:191854-sed}
\end{figure}

HD191854 is the clean case among the four: its two components have a 490~K temperature split (5930~K primary, 5440~K secondary), well separated relative to the precision of the data, and the fit converges without either component's temperature pinning against the CK04 grid's boundary (reduced $\chi^2=0.107$). Fig.~\ref{fig:191854-sed} shows this system's fitted composite SED against the literature photometry, chosen as this study's representative Al-Wardat SED fitting example.

HD214222, HD130669, and HD184467 each have a much smaller literature temperature split between components---75~K, 60~K, and 155~K, respectively---and in each case, one component's fitted temperature pins exactly at the local CK04 grid boundary. This was confirmed to be a genuine near-twin-temperature degeneracy rather than a grid-coverage artifact: extending the available grid range well past the literature temperature for both HD130669 and HD184467 does not resolve the pinning, it only relocates it, and the individual-component temperature match \emph{worsens} under the wider grid (HD130669's secondary-temperature error grew from 4.40\% to 13.14\%) while the combined-light $\chi^2$ barely changes either way. This is itself a meaningful result: an unresolved combined-light spectrum has very little power to separate two components' individual temperature once those temperatures are close to each other, however precise the photometry supplied, and reduced $\chi^2$ remaining excellent throughout is evidence that many different ($T_{\text{eff}, 1}$, $T_{\text{eff}, 2}$) splits reproduce the \emph{combined} light almost equally well when the two components are near-twins.

Str\"{o}mgren $u$ photometry, present in \citeauthor{Abushattal2025}'s Table 4, is excluded from all four fits: it disagrees with the composite model by approximately 1.7~mag, roughly a factor of 4.7 in flux---far outside the 0.25-0.5~mag systematic seen in every other band. Two independent zero-point sources were checked (the SVO Filter Profile Service's own Vega-relative zero point, and an independent CALSPEC-derived AB-to-Vega offset), and neither supports a correction of this size. No calibration adjustment was applied; Str\"{o}mgren $u$ remains excluded, since no independent source could confirm a correction of this size. Figures~\ref{fig:214222-sed}--\ref{fig:184467-sed} show these systems' fitted composite SEDs against the literature photometry.

%%%%%%%%%%%%%%%%%%%%%%%%%%%%%%%%%%%%%%%%%%%%%%%%%%%%%%%%%%%%%%
\subsection{Joint case studies}
\label{subsec:joint-val}
Capella and $\alpha$ Centauri AB are the two systems for which PySeshat's orbit-fitting and Al-Wardat SED-fitting capabilities are both validated for the same physical binary, using the joint inference layer of~Section\,\ref{subsec:mcmc}. For both systems, the fitted observations are synthetic---forward-modeled by PySeshat's own pipeline from independent sources at plausible measurement precisions, per the data-provenance note in~Section\,\ref{subsec:benchmarks}---rather than drawn from original archival data, since that data proved impractical to extract reliably within this study's scope. For clarity, these two systems validate whether the fitter recovers a published solution from synthetic data, which is meaningful but distinct in kind from~Section\,\ref{subsec:orbit-val}'s validation against real archival astrometry and RV data, and from~Section\,\ref{subsec:wardat-val}'s validation against an independent published photometric modeling output~(Section\,\ref{subsec:benchmarks}): here, both the fitted data and the fitting code originate from PySeshat itself, making this a self-consistency check on the forward and inverse models rather than a test of agreement with an external source.

\begin{figure*}
	\centering
	\includegraphics[width=\textwidth]{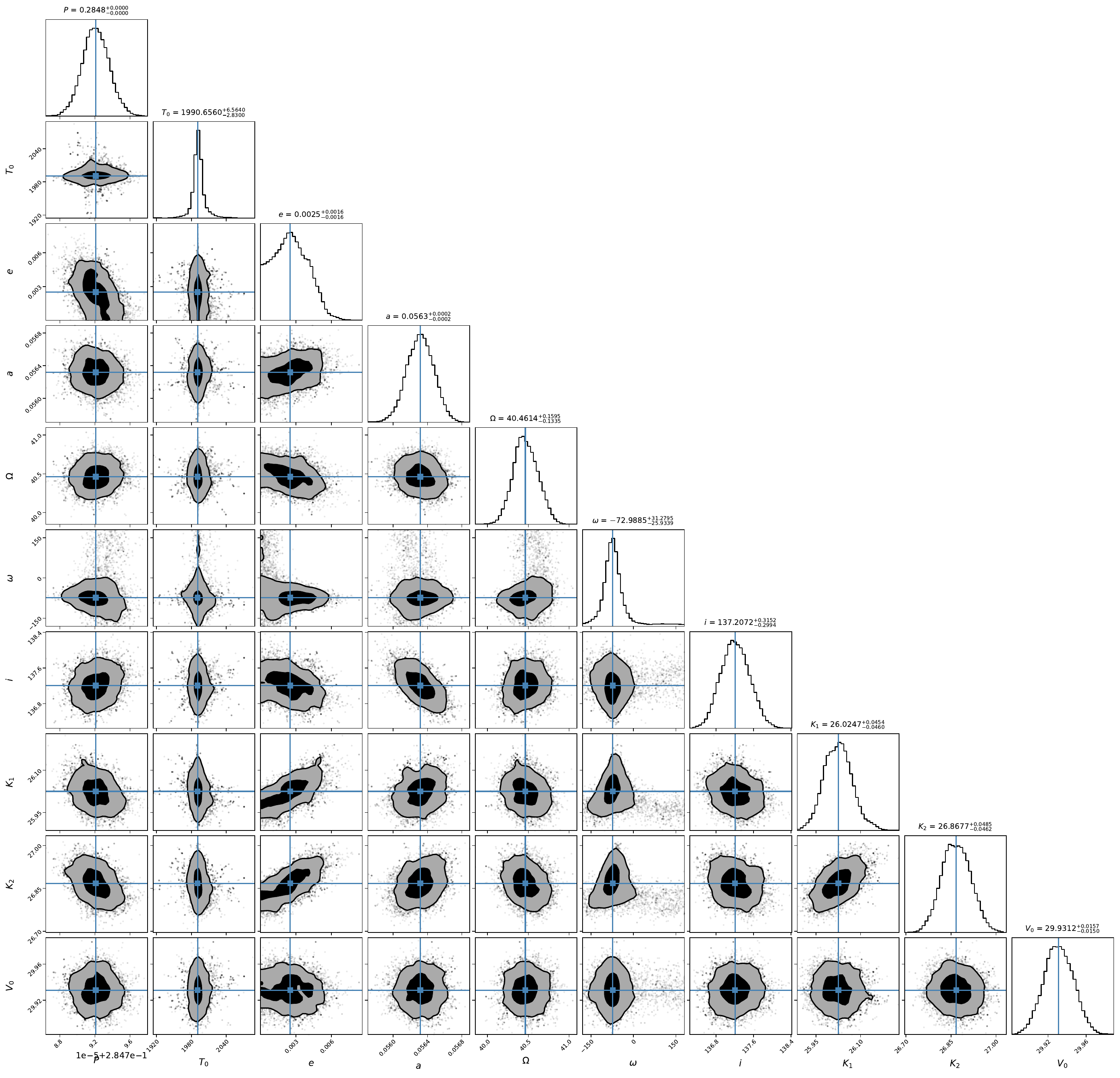}
	\caption{Posterior distributions from the MCMC fit of Capella's orbit, showing the characteristic near-circular-orbit degeneracy between eccentricity and argument of periastron: eccentricity is tightly constrained near zero (median $0.0025$, 68\% credible interval $[0.0009, 0.0041]$) while argument of periastron is broad (median $286.8^{\circ}$, 68\% credible interval $[260.7^{\circ}, 318.3^{\circ}]$; $\omega$ is plotted here on a $(-180^{\circ}, 180^{\circ}]$ scale for visual clarity, since its posterior straddles the $0^{\circ}/360^{\circ}$ wrap point under PySeshat's standard $[0^{\circ}, 360^{\circ})$ convention).}
	\label{fig:capella-cor}
\end{figure*}

Capella is a well-known, bright, evolved SB2 binary, fit against the orbital and stellar-parameter solution of \citet{Torres2015}, cross-checked against the independent SB2 solution of \citet{Weber2011}. Its orbit fit~(Fig.\,\ref{fig:capella-orbit}) recovers period, epoch of periastron, semi-major axis, node, inclination, both RV semi-amplitudes, and systemic velocity to $\le0.32$\% relative error (reduced $\chi^2=0.92$). Eccentricity and argument of  periastron are the sole exception, and an instructive one: Capella's literature eccentricity ($e=0.00089$) is very nearly circular, so PySeshat's fitted $e=0.0033$, while a large \emph{relative} error, is misleading take at face value. Decomposed into the physically well-conditioned combination $e\cos{\omega}$ and $e\sin{\omega}$, the fit recovers $e\cos{\omega}$ to $4.7\times10^{-5}$ absolute---essentially the astrometric noise floor---while $e\sin{\omega}$ is off by 0.0029: the orbit is tightly constrained along the line of apsides and poorly constrained perpendicular to it, the usual near-circular-orbit $e$/$\omega$ degeneracy. This is not a PySeshat-specific fitting weakness: an independent RadVel fit of the same data hits the identical degeneracy~(Section\,\ref{subsec:comp-val}), and it is directly visible in Capella's MCMC posterior. Fig.~\ref{fig:capella-cor} shows this posterior, chosen as this study's representative corner-plot example.

\begin{figure}
	\centering
	\includegraphics[width=\columnwidth]{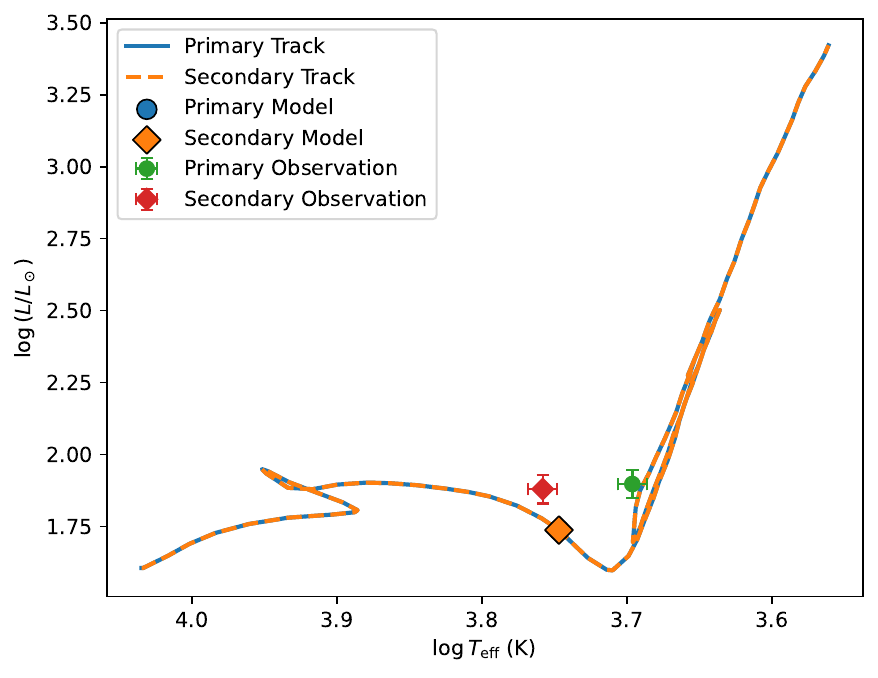}
	\caption{Capella's fitted components (points, with fitted uncertainties) placed on the \citet{Girardi2000} evolutionary tracks and isochrones~(Section\,\ref{subsec:plotting}), correctly recovering the system's known evolved status: the primary places as a giant and the secondary as a subgiant.}
	\label{fig:capella-hr}
\end{figure}

Capella's Al-Wardat SED fit recovers component radii to $\le4.24$\% and effective temperature to $\le3.55$\% of \citeauthor{Torres2015}'s values (reduced $\chi^2 = 2.12$)---comparable in quality to the four systems of~Section\,\ref{subsec:wardat-val}. Metallicity is pinned to the nearest available CK04 grid point (solar, $Z=0.019$) rather than Capella's published value ([Fe/H] $=-0.04$); this reflects the fitted value rounding to the nearest available metallicity slice. Capella's evolutionary consistency check~(Section\,\ref{subsec:plotting}) places its two components as a giant and a subgiant---correctly reflecting Capella's known evolved status, and a useful independent sanity check on the pipeline, since a system known from the literature to have evolved off the main sequence is not trivially guaranteed to land there in an HR diagram built from an independently fitted temperature and radius. Fig.~\ref{fig:capella-hr} shows this placement, chosen as this study's representative HR-diagram example.

$\alpha$ Centauri AB is a genuine Sun-like G2V+K1V SB2 pair with both a well-determined orbit and, via direct interferometry, among the most precisely known component temperatures and radii of any binary system. Orbital elements are taken from~\citet{Akeson2021}, cross-checked against~\citet{Pourbaix2016}; component effective temperatures from~\citet{Heiter2015}; and radii and surface gravities from~\citet{Kervella2017}. Neither source orbit paper includes individual RV semi-amplitudes for the two components---both use a combined astrometric-spectroscopic parameterization instead---so this system is fit using visual astrometry alone, an honest data gap rather than one filled with invented values.

The orbit fit~(Fig.\,\ref{fig:alpha-orbit}) recovers essentially exactly (reduced $\chi^2=1.05$, worst relative error $\le10^{-5}$\%), unsurprising given the system's well-separated, non-degenerate orbital geometry ($e\approx0.52$) and purely astrometric, noise-limited synthetic data. The Al-Wardat SED fit is a more instructive result: reduced $\chi^2=2.25$, with the secondary's fitted temperature the worst-recovered parameter at 14.7\% relative error. This was confirmed to be a genuine combined-light degeneracy local minimum rather than a grid-edge artifact. The fit is performed at solar metallicity despite the system's mildly super-solar literature value ([Fe/H] $\approx+0.23$): because $\alpha$ Centauri AB's photometry is itself synthetic, self-generated at solar metallicity~(Section\,\ref{subsec:benchmarks}), fitting at solar is the self-consistent choice for this system rather than a grid limitation---refitting at the literature-closer metallicity value was tested and produced a worse fit (reduced $\chi^2$ from $2.25$ to $2.92$), confirming the synthetic data's own assumed composition rather than indicating an error. Unlike Capella, $\alpha$ Centauri AB's evolutionary consistency check places both components as ordinary main-sequence dwarfs, correctly reflecting their known near-solar, unevolved status~(Fig.\,\ref{fig:alpha-hr}).

%%%%%%%%%%%%%%%%%%%%%%%%%%%%%%%%%%%%%%%%%%%%%%%%%%%%%%%%%%%%%%
\section{Discussion}
\label{sec:disc}
The clearest methodological result of this validation is not any single benchmark's numerical agreement, but the direct empirical demonstration, across two independently structured tests, of why simultaneous inner-and-outer orbit fitting is necessary for hierarchical triples rather than a refinement of secondary importance. $\zeta$ Aquarii shows this constructively: PySeshat's simultaneous ORBIT3 fit improves on a previously published solution, and its legacy-formalism counterpart converges to a consistent, independently-derived result via an entirely different parameterization. HIP\,111805 shows the same point destructively: a formalism that cannot represent this system's real orbital configuration at all does not fail gracefully when forced to try, it drives itself into a numerically singular fit. Together, these two systems make a stronger case for the modern formalism than either could alone---one demonstrates correctness, the other demonstrates necessity.

The photometric side of this validation carries a comparable methodological result, and one that generalizes beyond the four systems it was observed in: unresolved-photometry SED fitting cannot reliably separate two components' individual temperatures once those temperatures are close to each other, independent of how many photometric bands are available or how precisely they are measured. This is not a limitation specific to PySeshat's implementation of the Al-Wardat method; it is a property of the underlying inverse problem, confirmed here by the fact that widening the available model grid relocates the degeneracy rather than resolving it. This finding is, in effect, an independent rediscovery of the reason the Al-Wardat method has historically been paired with evolutionary-track consistency checks rather than deployed on photometry alone---a methodological choice in the literature that this validation now has a first-principles justification for, rather than treating it as an inherited convention.

A related, smaller-scale instance of the same underlying issue arose in the evolutionary consistency checks themselves, beyond what~Section\,\ref{sec:validation} reported directly: for at least one Al-Wardat benchmark system, a fitted component's position falls close enough to an evolutionary-track grid boundary that its individual age and mass are only weakly constrained in isolation, with the ambiguity resolved only by fitting both components' ages jointly rather than trusting either independently. This is the same class of degeneracy as the temperature-splitting result above, arising in a different part of the pipeline: a single star's position in an observable space is often compatible with a range of underlying physical states, and robustness in this method comes from combining multiple stars' or multiple data types' constraints rather than from the precision of any one measurement alone.

The provenance of the comparison photometry used in~Section\,\ref{subsec:wardat-val} deserves a second, explicit mention here because it changes how the SED-fitting validation should be read as a whole. Because \citeauthor{Abushattal2025}'s Table~4 is itself synthetic output from an independent modeling pipeline rather than raw catalog photometry, the four HD-system benchmarks demonstrate agreement between two independent implementations of comparable synthetic-photometry pipelines evaluated against the same underlying stellar parameters---not agreement between a physical model and a direct sky measurement. This is a meaningful and non-trivial validation, since it rules out implementation-specific error in PySeshat's own SED code relative to a second, independent codebase. It is not, however, evidence that either pipeline's synthetic photometry matches what a telescope would actually record for these systems, a distinct claim this study does not make.

Positioned against the broader software landscape described in~Section\,\ref{sec:intro}, PySeshat's contribution is less about outperforming any single existing tool on its own ground and more about occupying a gap between tools that each solve one part of this problem well. \texttt{orbitize!} and Gaia's automated pipeline both address visual-orbit determination competently and at scales PySeshat does not attempt to match, but neither extends to hierarchical triples or stellar atmosphere modeling; RadVel and PyAstronomy serve adjacent, narrower roles as the independent-package comparisons in~Section\,\ref{subsec:comp-val} make explicit; and the Al-Wardat method itself has, to date, existed as a methodology applied through bespoke, system-specific analysis rather than as a general, open, testable software package. PySeshat's value is in unifying these capabilities behind a common, regression-tested interface, and the negative results in this validation---the legacy triple formalism's failure mode, the temperature-splitting degeneracy---are as much a part of that contribution as the successful benchmark reproductions, since they mark out where boundaries of what any implementation of these methods can and cannot resolve actually lie.

Two limitations from~Section\,\ref{sec:validation} are worth restating here specifically because they bound how the results above should be generalized, rather than as new findings. The Str\"{o}mgren $u$ exclusion~(Section\,\ref{subsec:wardat-val}) means this study's SED-fitting validation is not tested against that passband's contribution to the fit; whether it would sharpen or degrade the temperature-splitting result in future work with a correctly calibrated zero point is an open question. And the joint case studies~(Section\,\ref{subsec:joint-val}) validate the fitter's ability to recover a known solution from synthetic data, which is a necessary but not sufficient condition for trusting the same pipeline on a real, previously unpublished system's raw archival data---the latter remains, for these two specific systems, future work rather than something this study demonstrates directly.

%%%%%%%%%%%%%%%%%%%%%%%%%%%%%%%%%%%%%%%%%%%%%%%%%%%%%%%%%%%%%%
\section{Conclusions}
\label{sec:conc}
We have presented PySeshat, an open-source Python package unifying classical and Bayesian binary and hierarchical-triple orbit determination with Al-Wardat-style SED fitting in a single, tested pipeline---a combination of capabilities not previously available together in one open tool. The software was validated end-to-end against twelve real, published benchmark systems: six through orbit fitting, including two hierarchical triples fit through the full simultaneous inner-and-outer ORBIT3 formalism; four through Al-Wardat SED fitting against a paper coauthored by Al-Wardat himself; and two---Capella and $\alpha$ Centauri AB---through both capabilities jointly for the same physical system.

Beyond reproducing published solutions, this validation produced two results with implications beyond PySeshat's own implementation. First, direct comparison between PySeshat's modern and legacy hierarchical-triple formalisms on the same real systems demonstrates not merely that the modern ORBIT3 formalism is preferable, but that it is necessary: a real, published system's orbital configuration was shown to exceed what the legacy formalism can represent at all, independent of fit quality. Second, unresolved-photometry SED fitting was shown to be unable to reliably separate two components' individual temperatures once those temperatures are close to each other, regardless of photometric band count or model-grid coverage---a finding that reproduces, from first principles, the methodological rationale for pairing SED fitting with independent evolutionary-track constraints in the Al-Wardat method's own established practice.

The software's own orbit-fitting core was additionally cross-validated against three independent packages (RadVel, PyAstronomy, \texttt{orbitize!}) and its Gaia DR3 NSS parser against real archive-retrieved data for two systems, in each case confirming internal correctness against implementations outside PySeshat's own codebase. Limitations disclosed throughout this work---the synthetic provenance of the comparison photometry in~Section\,\ref{subsec:wardat-val}, the self-consistency-only nature of the two joint case studies in~Section\,\ref{subsec:joint-val}, and the excluded Str\"{o}mgren $u$ passband---are stated explicitly so that the scope of what has and has not been demonstrated is clear to future users of the package.

PySeshat is released as open-source software, and is installable via the PyPI (Python Package Index). We intend it to serve both as a validated tool for binary and multiple-star systems analysis and as a foundation that can be extended in future work, particularly toward real archival validation of the two currently synthetic joint case studies, broader stellar-atmosphere and evolutionary-track grid coverage, and a graphical interface (GUI) to lower the barrier to entry for users less familiar with Python.
%%%%%%%%%%%%%%%%%%%%%%%%%%%%%%%%%%%%%%%%%%%%%%%%%%%%%%%%%%%%%%
\begin{acknowledgements}
	This research made use of NumPy~\citep{Harris2020}, SciPy~\citep{Virtanen2020}, and Matplotlib~\citep{Hunter2007}, for numerical computation and visualization; \texttt{emcee}~\citep{ForemanMackey2013} and \texttt{corner}~\citep{ForemanMackey2016} for Bayesian posterior sampling and visualization; and \texttt{adjustText} (\url{https://github.com/Phlya/adjustText}) for automated label placement.
	
	We also acknowledge the independent software packages used for cross-validation: RadVel~\citep{Fulton2018}, PyAstronomy~\citep{Czesla2019}, and \texttt{orbitize!}~\citep{Blunt2020}.
	
	This work has made use of data from the European Space Agency (ESA) mission \emph{Gaia} (\url{https://www.cosmos.esa.int/gaia}), processed by the Gaia Data Processing and Analysis Consortium (DPAC, \url{https://www.cosmos.esa.int/web/gaia/dpac/consortium}). Funding for the DPAC has been provided by national institutions, in particular the institutions participating in the Gaia Multilateral Agreement. Gaia DR3 data used were retrieved via the Gaia Archive web interface.
	
	This research has made use of the VizieR catalog access tool, CDS, Strasbourg, France~\citep{Ochsenbein1996}. The original description of the VizieR service was published in Ochsenbein, Bauer \& Marcout~\citep{Ochsenbein2000}. The \citet{Girardi2000} evolutionary tracks and isochrones were retrieved from VizieR catalog J/A+AS/141/371.
	
	This research has made use of the SIMBAD database, operated at CDS, Strasbourg, France.
	
	This work made use of Vega-system zero points from the SVO Filter Profile Service~\citep{Rodrigo2020, Rodrigo2012}.
	
	This work made use of the \citet{Castelli2004} grid of model stellar atmospheres.
	
	PySeshat's binary and hierarchical-triple orbit-fitting formalisms follow, and are validated against, Tokovinin's IDL-based ORBIT and ORBIT3 software~\citep{Tokovinin2016, Tokovinin2017}.
\end{acknowledgements}

\bibliographystyle{aa} % style aa.bst
\bibliography{refs} % your references Yourfile.bib

\onecolumn
\begin{appendix}
	\onecolumn
	\section{Additional figures}
	\label{app}
	\begin{figure}[ht!]
		\centering
		\begin{subfigure}{\columnwidth}
			\centering
			\includegraphics[height=0.29\textheight]{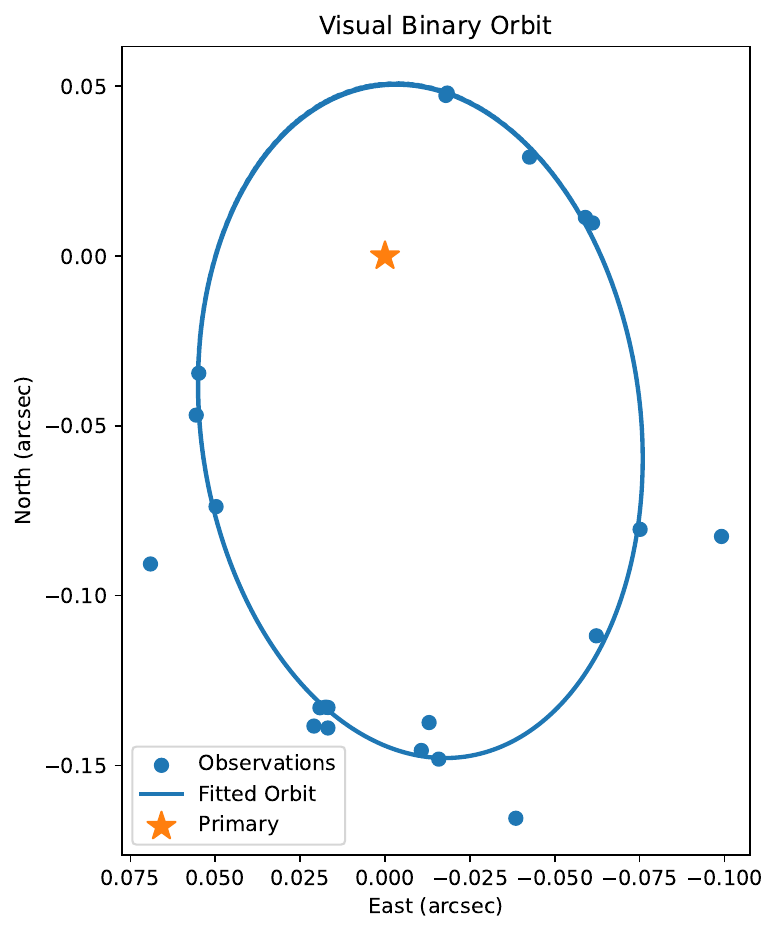}
			\subcaption{~}
			\label{fig:fin_panel_a}
		\end{subfigure}
		
		\vspace{2mm}
		
		\begin{subfigure}{\columnwidth}
			\centering
			\includegraphics[height=0.26\textheight]{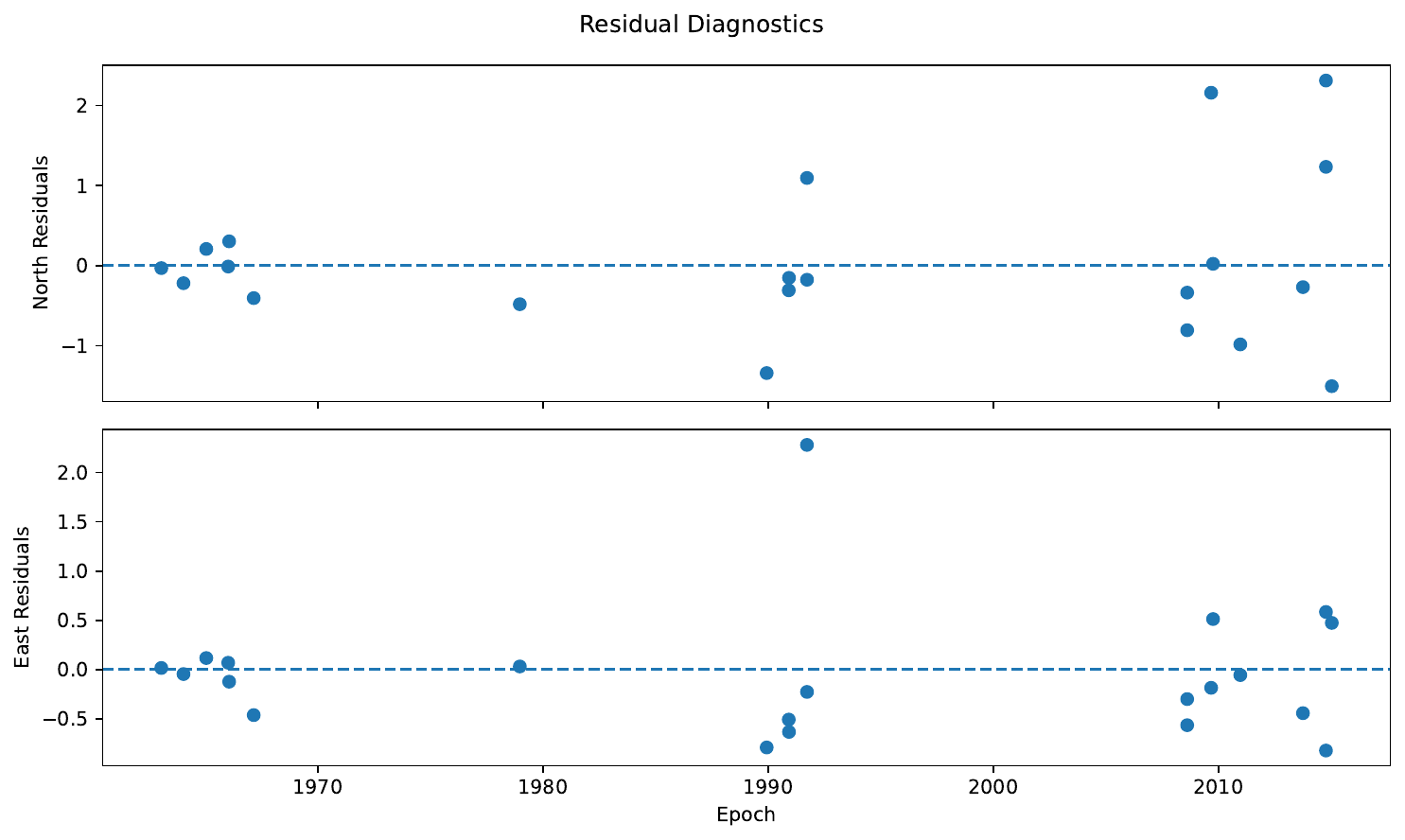}
			\subcaption{~}
			\label{fig:fin_panel_b}
		\end{subfigure}	
		
		\vspace{2mm}
		
		\begin{subfigure}{\columnwidth}	
			\centering
			\includegraphics[height=0.26\textheight]{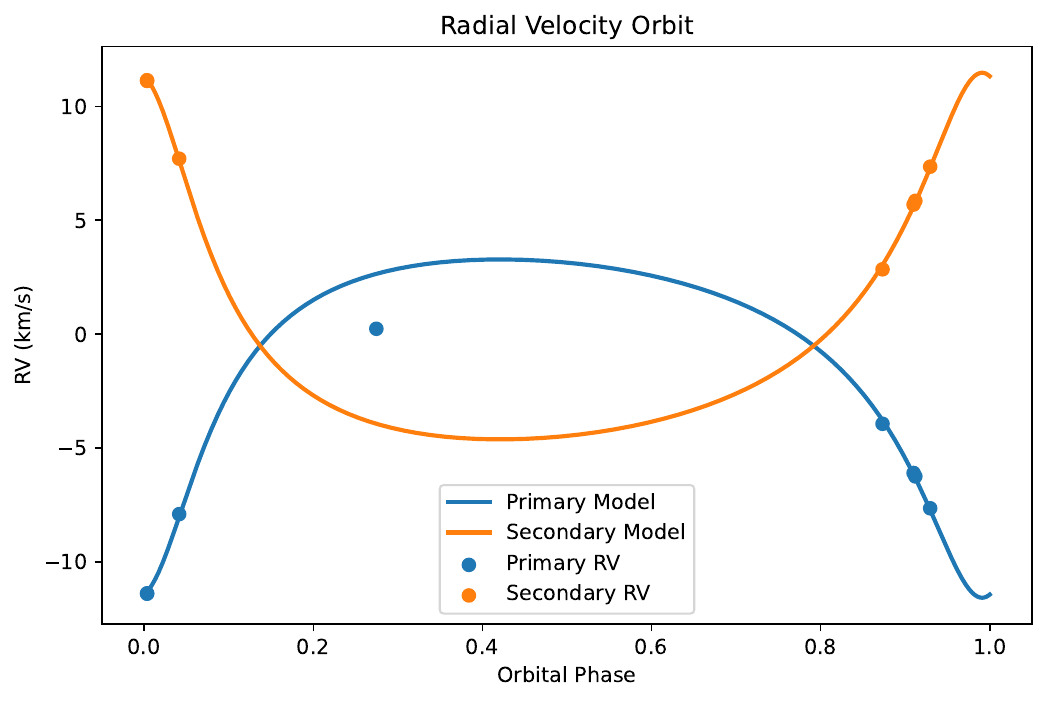}
			\subcaption{~}
			\label{fig:fin_panel_c}
		\end{subfigure}
		\caption{FIN379 orbit-fitting results. (a) Fitted visual orbit. (b) Observed-minus-calculated (O--C) residuals in Cartesian coordinates. (c) Phase-folded radial-velocity curves for both components.}
		\label{fig:app-fin}
	\end{figure}
	
	\begin{figure}
		\centering
		\includegraphics[height=0.48\textheight]{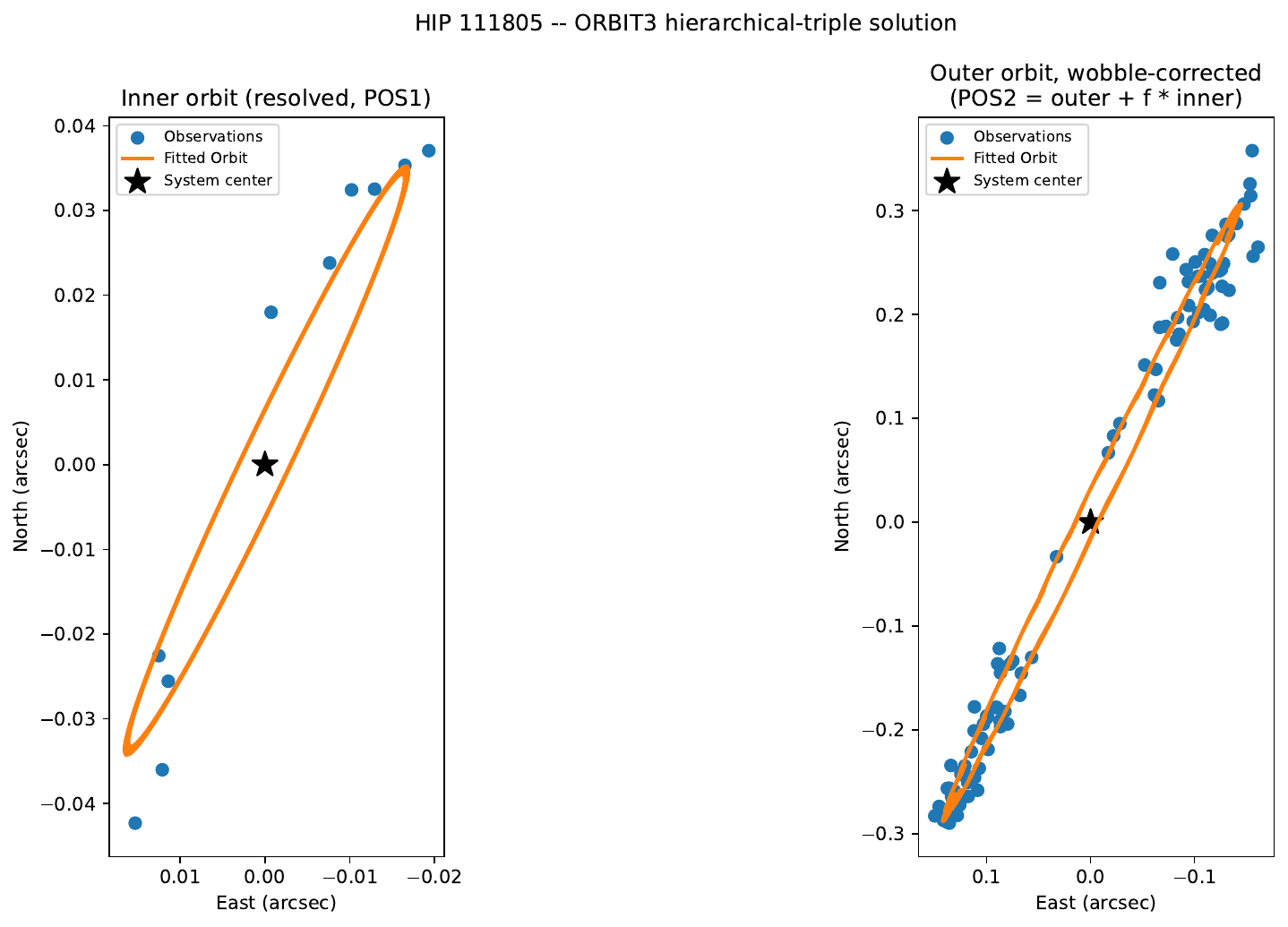}
		\caption{HIP\,111805 ORBIT3 fit. \emph{Left panel}: Fitted inner orbit with observed inner-pair astrometric positions. \emph{Right panel}: Fitted wobble-corrected outer orbit with observed outer-pair astrometric positions.}
		\label{fig:app-111805}
	\end{figure}
	
	\begin{figure}
		\centering
		\includegraphics[height=0.42\textheight]{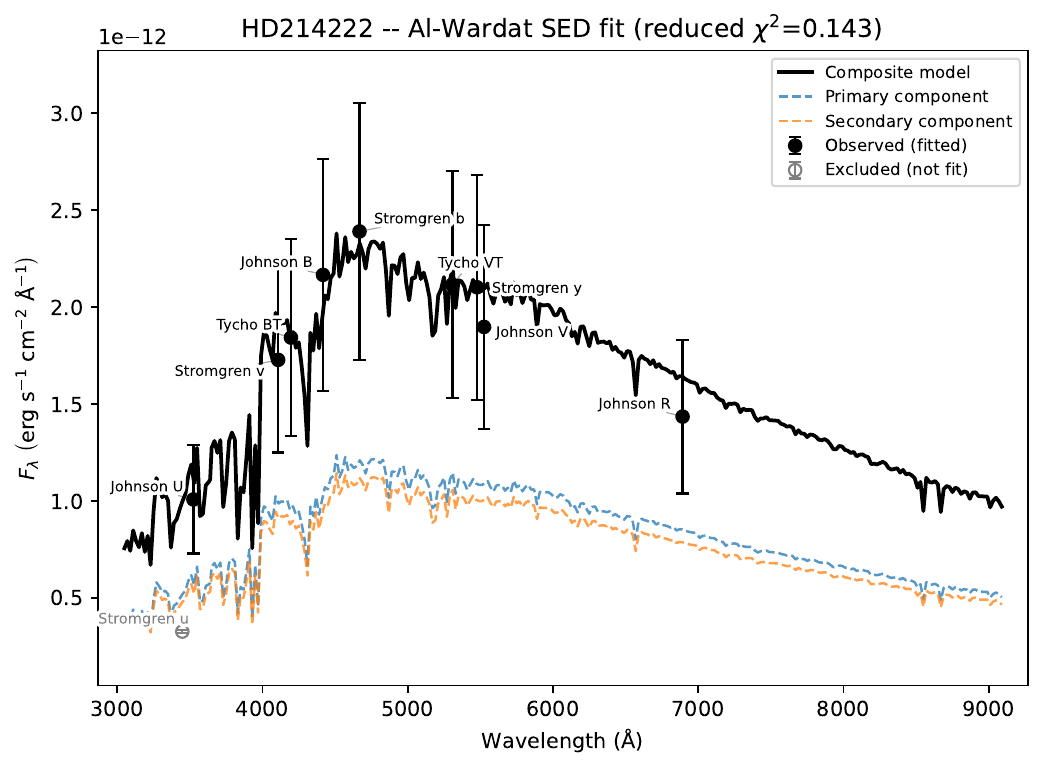}
		\caption{HD214222 Al-Wardat SED fit.}
		\label{fig:214222-sed}
	\end{figure}
	
	\begin{figure}
		\centering
		\includegraphics[height=0.42\textheight]{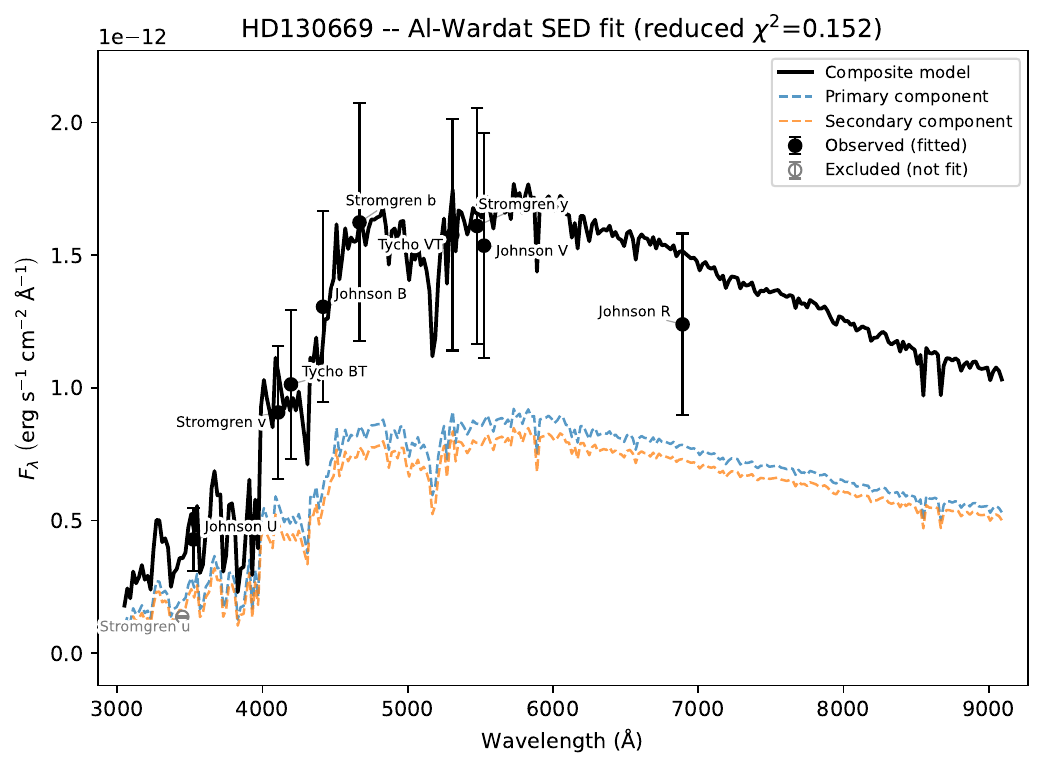}
		\caption{HD130669 Al-Wardat SED fit.}
		\label{fig:130669-sed}
	\end{figure}
	
	\begin{figure}
		\centering
		\includegraphics[height=0.42\textheight]{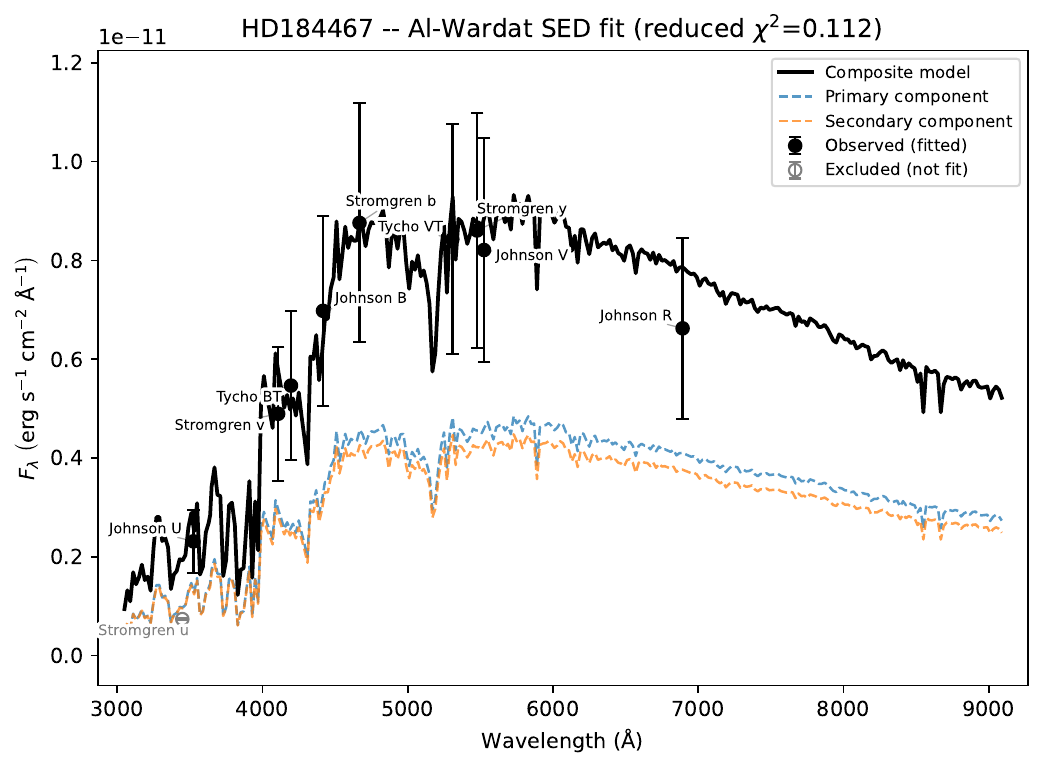}
		\caption{HD184467 Al-Wardat SED fit.}
		\label{fig:184467-sed}
	\end{figure}
	
	\begin{figure}[ht!]
		\centering
		\begin{subfigure}{\columnwidth}
			\centering
			\includegraphics[height=0.29\textheight]{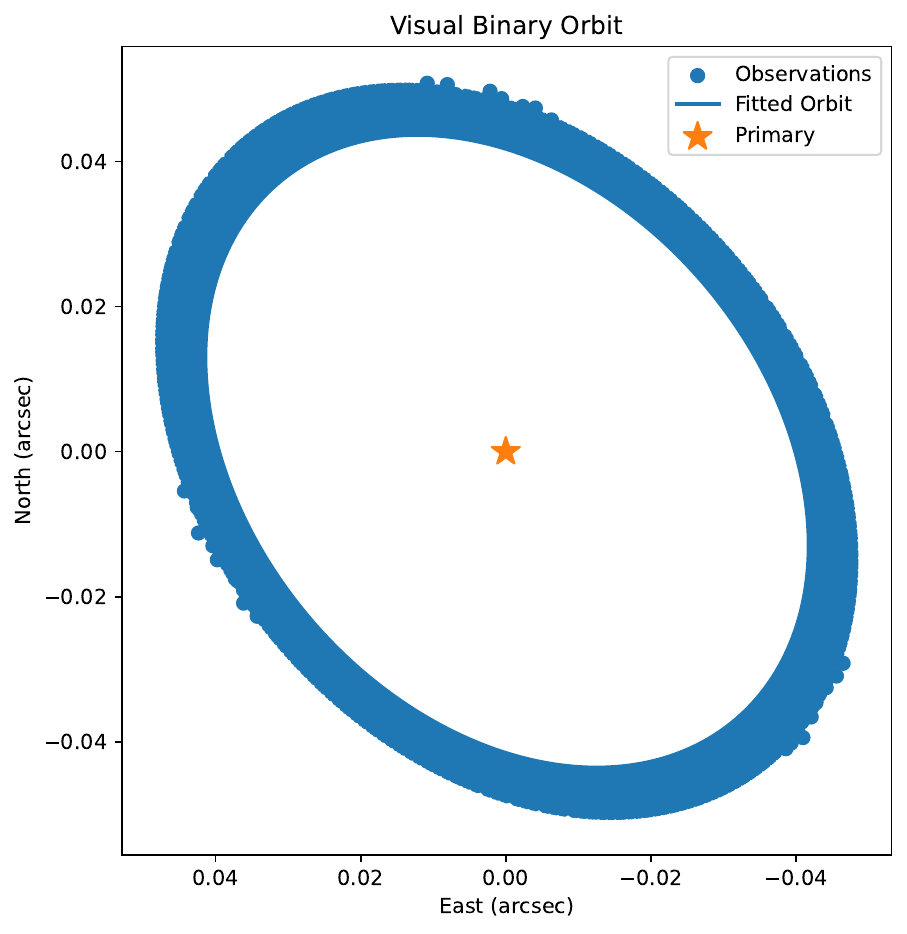}
			\subcaption{~}
			\label{fig:capella_panel_a}
		\end{subfigure}
		
		\vspace{2mm}
		
		\begin{subfigure}{\columnwidth}
			\centering
			\includegraphics[height=0.26\textheight]{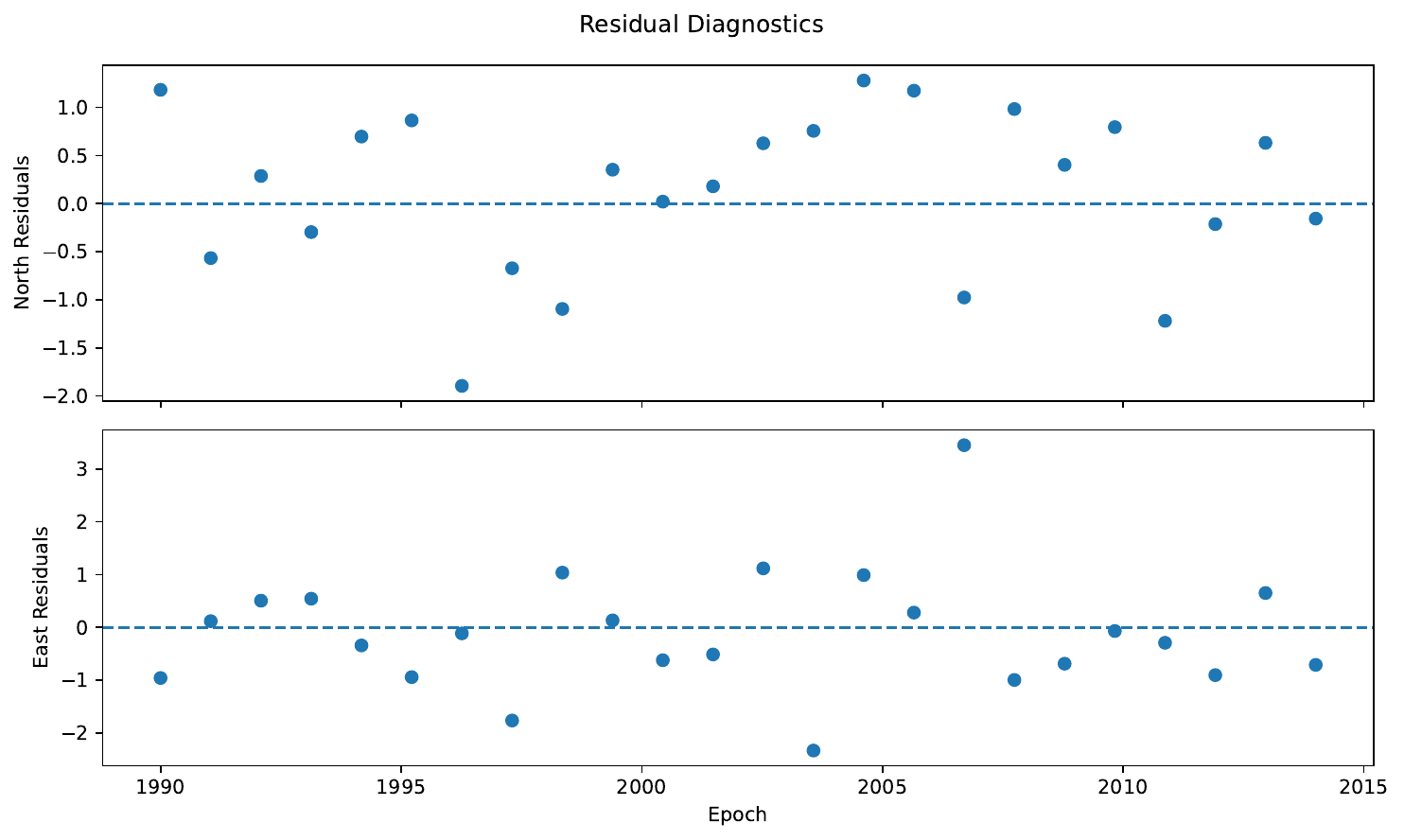}
			\subcaption{~}
			\label{fig:capella_panel_b}
		\end{subfigure}	
		
		\vspace{2mm}
		
		\begin{subfigure}{\columnwidth}	
			\centering
			\includegraphics[height=0.26\textheight]{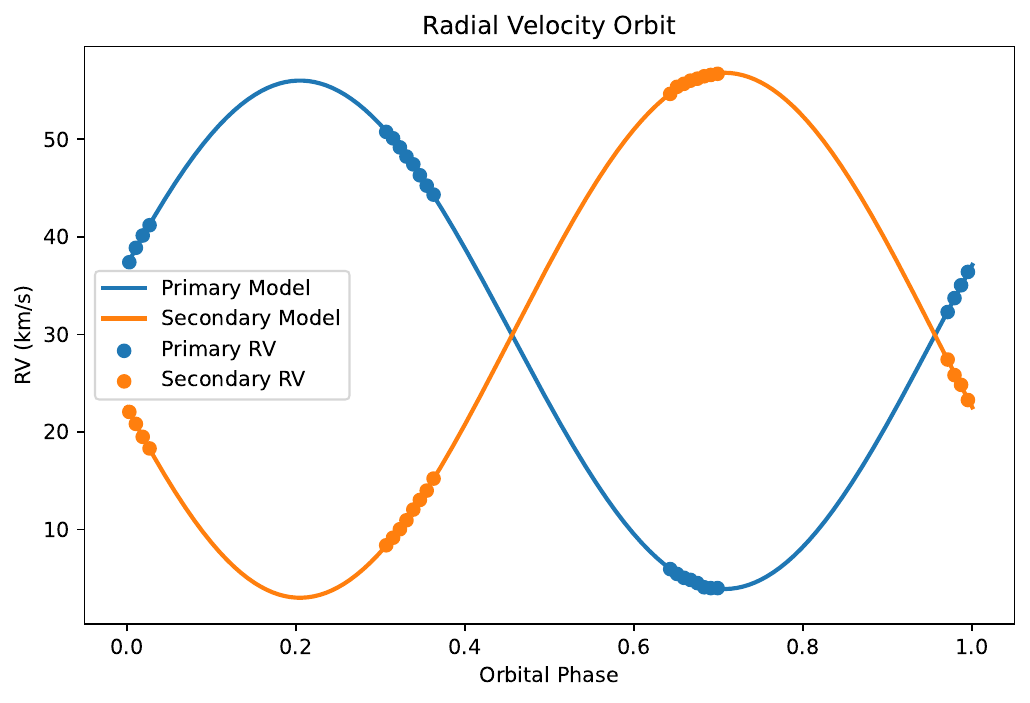}
			\subcaption{~}
			\label{fig:capella_panel_c}
		\end{subfigure}
		\caption{Capella orbit-fitting results. (a) Fitted visual orbit. (b) Observed-minus-calculated (O--C) residuals in Cartesian coordinates. (c) Phase-folded radial-velocity curves for both components.}
		\label{fig:capella-orbit}
	\end{figure}
	
	\begin{figure}
		\centering
		\begin{subfigure}{\columnwidth}
			\centering
			\includegraphics[height=0.29\textheight]{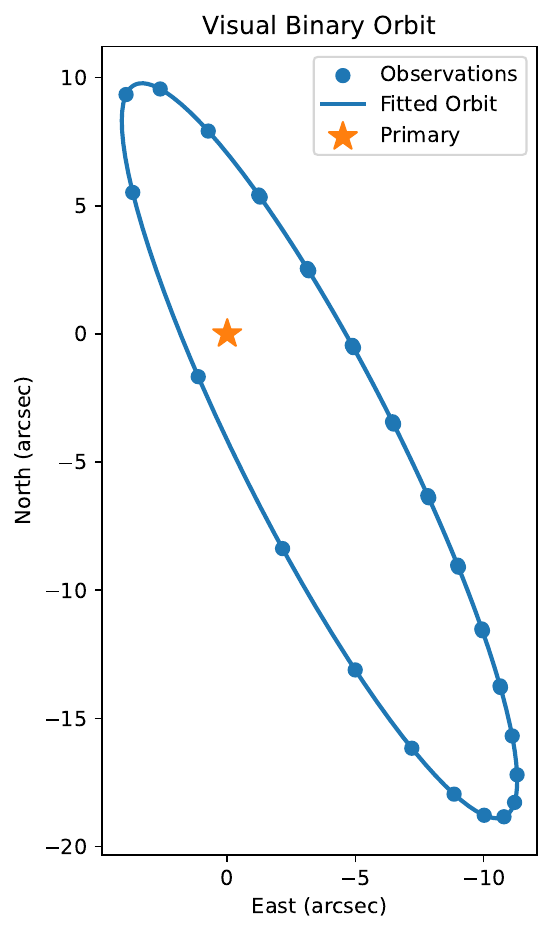}
			\caption{}
			\label{fig:alpha_panel_a}
		\end{subfigure}
		\vspace{2mm}
		\begin{subfigure}{\columnwidth}
			\centering
			\includegraphics[height=0.26\textheight]{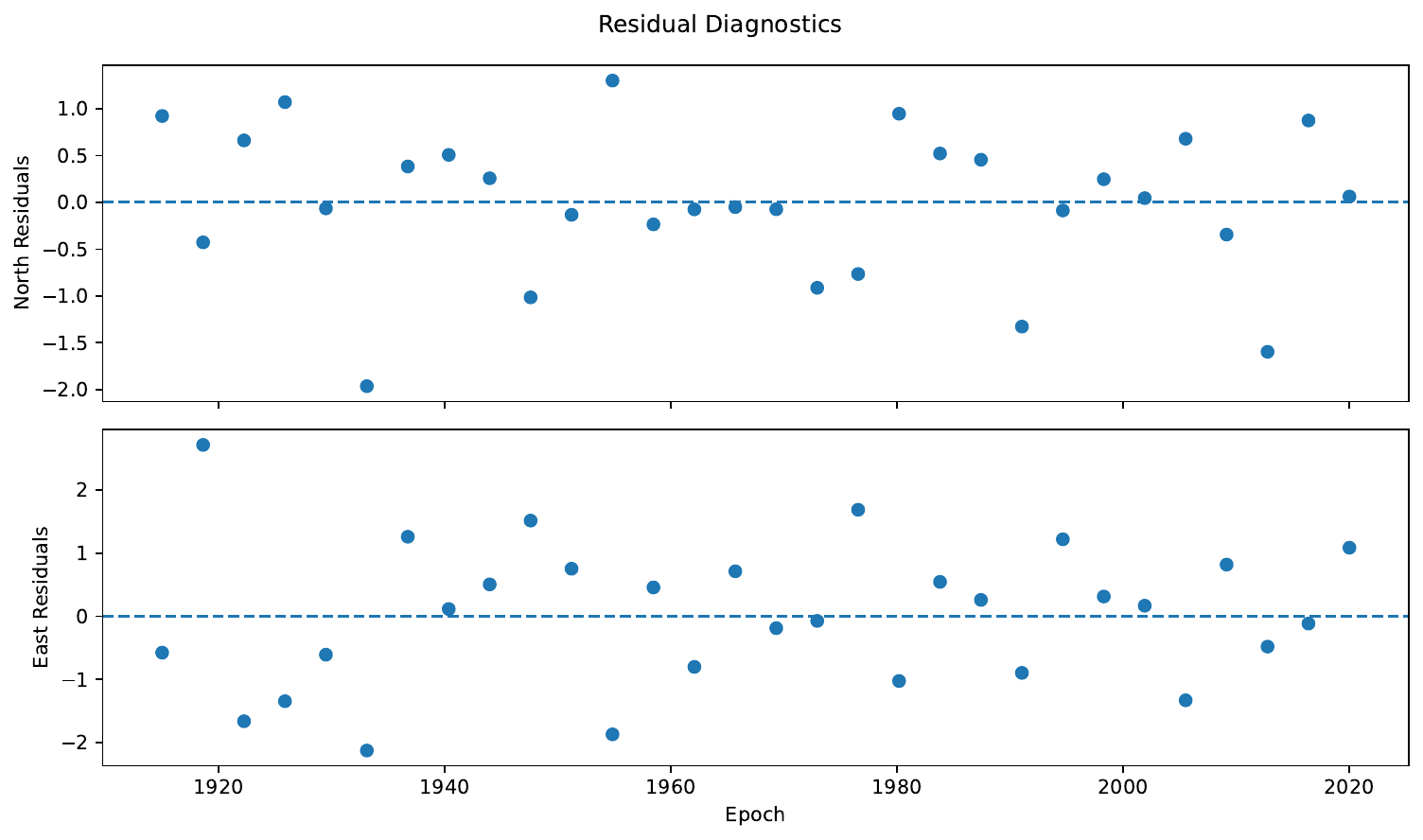}
			\caption{}
			\label{fig:alpha_panel_b}
		\end{subfigure}
		\caption{$\alpha$ Centauri AB orbit-fitting results. (a) Fitted visual orbit. (b) Observed-minus-calculated (O--C) residuals in Cartesian coordinates.}
		\label{fig:alpha-orbit}
	\end{figure}

	\begin{figure}
		\centering
		\includegraphics[height=0.42\textheight]{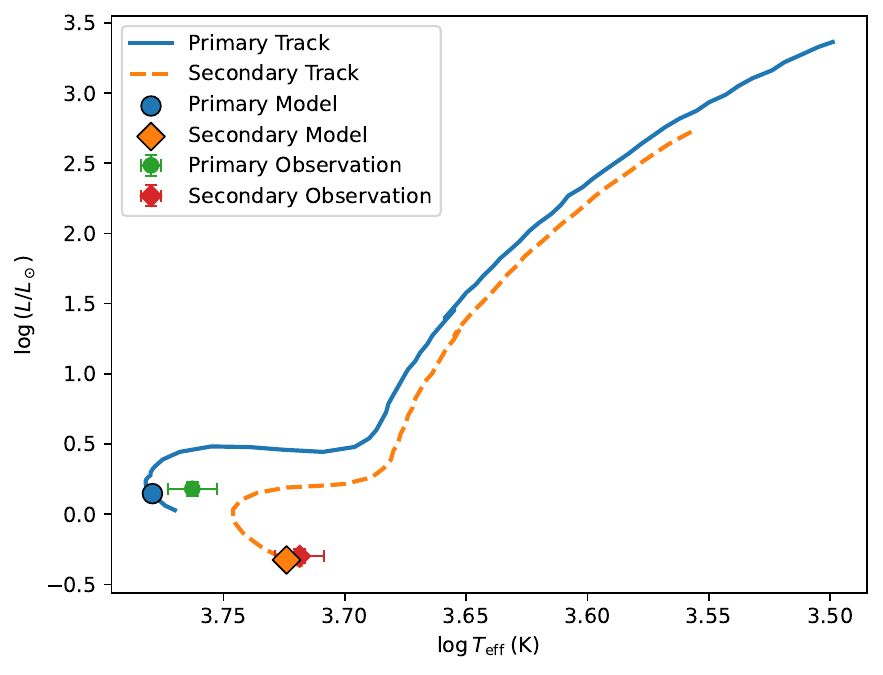}
		\caption{$\alpha$ Centauri AB fitted components placed on the evolutionary tracks and isochrones. Both components place as main-sequence stars.}
		\label{fig:alpha-hr}
	\end{figure}

\end{appendix}
\end{document}